\documentclass[5p,a4paper,final,twocolumn]{elsarticle}

\usepackage{footnote}
\newenvironment{alphafootnotes}
  {\par\edef\savedfootnotenumber{\number\value{footnote}}
   
   \setcounter{footnote}{0}}
  {\par\setcounter{footnote}{\savedfootnotenumber}}

\usepackage{graphics}
\usepackage{lineno}
\usepackage{subcaption}	%the new standard for subfloats
\usepackage{amsmath,amssymb,amsfonts,amsthm,amsbsy,amstext,stmaryrd}
\usepackage{siunitx}		% allows the correct half-space when typing units,e.g.
\usepackage{xfrac}
\usepackage[super]{nth}

\usepackage{booktabs} 	% extra table features, like \toprule and vertical rules
\usepackage{multirow} 	% use multirow and multicolumn cells in tables

\usepackage{url}

\newcommand{\etal}{\emph{et al.}\ }

\newcommand{\Fe}{\text{Fe}}
\newcommand{\Al}{\text{Al}}
\newcommand{\Be}{\text{Be}}
\newcommand{\FeBe}{\text{FeBe}}
\newcommand{\AlFeBe}{\text{AlFeBe}}
\newcommand{\s}{(\text{s})}

\biboptions{square,comma,sort&compress}

\journal{J.~Alloy.~Compd.}

\begin{document}

\begin{frontmatter}

%%% PRB Title
\title{Crystal structure, thermodynamics, magnetics and disorder properties of Be-Fe-Al intermetallics}
%%% General Title
%\title{Stability and structures of Be-Fe-Al intermetallics.}

\author[IC,ANSTO]{P.A. Burr}
\author[ANSTO]{S.C. Middleburgh\corref{corresponding}} \ead{simm@ansto.gov.au}
\author[IC]{R.W. Grimes}
\address[IC]{Centre for Nuclear Engineering and Department of Materials, Imperial College London, London, SW7 2AZ, UK.}
\address[ANSTO]{Institute of Materials Engineering, Australian Nuclear Science \& Technology Organisation, Menai, New South Wales 2234, Australia.}
\cortext[corresponding]{Corresponding author.}

%%	General abstract
%\begin{abstract}
%Phases that are relevant to Be alloys with low concentrations of Al and Fe are investigated using \emph{ab-initio} atomic scale simulations, coupled with phonon density of states calculations to capture temperature effects. The elastic and magnetic properties, thermodynamical stability, deviation from stoichiometry and order/disorder transformations were investigated for the intermetallics FeBe$_2$, FeBe$_5$, the $\varepsilon$ phase (predicted to be Fe$_{2+x}$Be$_{17-x}$) and AlFeBe$_4$.
%In absence of Al, FeBe$_5$ is predicted to form at equilibrium above ${\sim}1250$~K, while the $\varepsilon$ phase is stable only below ${\sim}1650$~K, and FeBe$_2$ is stable at all temperatures below melting.
%Small additions of Al are found to stabilise FeBe$_5$ over FeBe$_2$ and $\varepsilon$, while at high Al content, AlFeBe$_4$ is predicted to form.
%Deviations from stoichiometric compositions are also considered and found to be important in the case of FeBe$_5$ and $\varepsilon$. The propensity for disordered vs ordered structures is also important for AlFeBe$_4$ (which exhibits complete Al-Fe disordered at all temperatures) and FeBe$_5$ (which exhibits an order-disorder transition at ${\sim}950$~K).
%\end{abstract}

%%	Phys Rev B abstract
\begin{abstract}
The elastic and magnetic properties, thermodynamical stability, deviation from stoichiometry and order/disorder transformations of phases that are relevant to Be alloys were investigated using density functional theory simulations coupled with phonon density of states calculations to capture temperature effects.
A novel structure and composition were identified for the Be-Fe binary $\varepsilon$ phase.
In absence of Al, FeBe$_5$ is predicted to form at equilibrium above ${\sim}1100$~K, while the $\varepsilon$ phase is stable only below ${\sim}1500$~K, and FeBe$_2$ is stable at all temperatures below melting.
Small additions of Al are found to stabilise FeBe$_5$ over FeBe$_2$ and $\varepsilon$, while at high Al content, AlFeBe$_4$ is predicted to form.
Deviations from stoichiometric compositions are also considered and found to be important in the case of FeBe$_5$ and $\varepsilon$. The propensity for disordered vs ordered structures is also important for AlFeBe$_4$ (which exhibits complete Al-Fe disordered at all temperatures) and FeBe$_5$ (which exhibits an order-disorder transition at ${\sim}950$~K).
\end{abstract}

\end{frontmatter}

%\linenumbers

%%%%%%%%%%%%%%%%%%%%%%%%%%%%%% Main Text %%%%%%%%%%%%%%%%%%%%%%%%%%%%

%%%%%%%%%%%%%%%       INTRO       %%%%%%%%%%%%%%%%
\section{Introduction}
\label{sec:intro}
Beryllium (Be) is a light element with excellent neutron transparency, and for this reason it is
currently used as a plasma facing material in the Joint European Torus (JET) fusion reactor \cite{Deksnis1997} and has been selected for use in the International Thermonuclear Experimental Reactor (ITER) \cite{Thompson2007}.
However, Be is also highly toxic \cite{Kriebel1988}, which makes experimental research and development of Be alloys a difficult and expensive task.
In the extreme environment associated with the fusion plasma, the presence of impurities and alloying additions may play a crucial role in the ageing and degradation processes.
If the impurity elements are not retained in solution within the Be phase, they will form second phase particles embedded within the grains or form at grain boundaries and surfaces, where their presence can be deleterious to the mechanical and chemical properties of the alloy.
Here we will be concerned with the iron (Fe) and aluminium (Al) containing intermetallic phases of Be, as Fe and Al are common additions/impurities in Be alloys \cite{Rooksby1962a}.

\begin{figure}[hbt]
      \centering
      \includegraphics[width=0.5\textwidth]{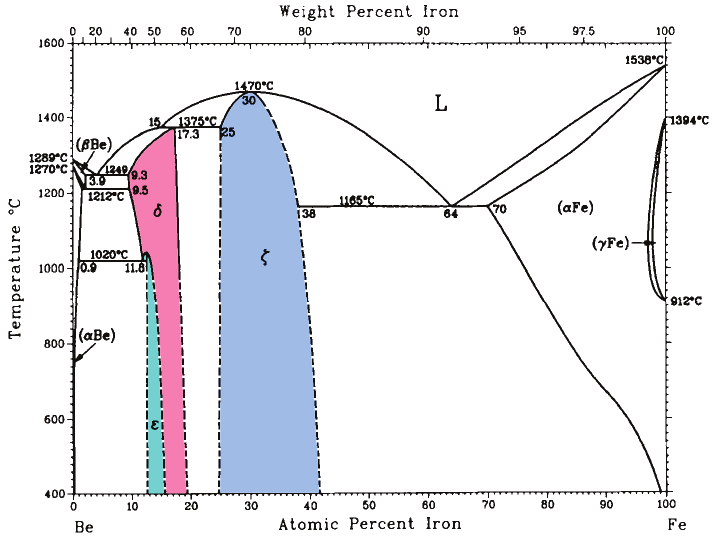}
      \caption{Be-Fe phase diagram reproduced from \cite{Okamoto1988}, with intermetallic phases highlighted in colour.}
      \label{fig:phaseDiagram}
\end{figure}

In a review of the binary Be-Fe system, Tanner and Okamoto \cite{Okamoto1988} highlight that, despite the many conflicting reports, much of the phase diagram is now well characterised, see fig.~\ref{fig:phaseDiagram}. However, `more data on the thermodynamic properties and phase diagram of this system are needed to improve the model' \cite{Okamoto1988}. This system exhibits solid solutions at either end of the composition range, a metastable BeFe$_3$ phase and three stable intermetallic compounds: FeBe$_2$ ($\zeta$), FeBe$_5$ ($\delta$) and a Be-rich intermetallic phase ($\varepsilon$) of unknown structure and uncertain composition.
A recent {\sc claphad} study \cite{Ohtani2004} expanded the understanding of Be solution in $\alpha$-Fe, by including the effect of magnetic transition and order-disorder transitions. They also propose that the $\delta$ phase undergoes a first order transition at $\sim 1150$ K, decomposing into $\varepsilon$ and $\zeta$ below said temperature. This was not captured in earlier computational work \cite{Kaufman1984}, and experimental investigation was chiefly concerned with temperatures above $\sim 1200$ K \cite{Okamoto1988}.

Regarding the $\varepsilon$ phase,
Teitel and Cohen \cite{Teitel1948} first report it as hexagonal with composition FeBe$_{11}$ (or potentially FeBe$_{12}$), which forms only at temperatures below \SI{1065}{\celsius} and exhibits limited solubility (7.8--8.2 at.~\% Fe). By means of density measurements, their work shows that a unit cell of FeBe$_{11}$ should contain 18 atoms (i.e. $1\sfrac{1}{2}$ formula units), which remains a peculiar and unexplained result.
In subsequent publications, this hexagonal phase is the most commonly reported \cite{Okamoto1988}, however there is conflicting information. For example, Von Batchelder and Raeuchle \cite{VonBatchelder1957} proposed a body centred tetragonal Mn$_{12}$Th-type structure. Hindle and Slattery \cite{Hindle1963} reports a body-centred tetragonal Be-rich compound. Johnson \etal \cite{Johnson1970} report a new hexagonal phase with unknown composition FeBe$_x$, but assign the space group $P\bar{6}m2$ and a basis consisting of 19 symmetrically unique sites, though potentially with partial occupancy. X-ray diffraction reveals many similarities with the FeBe$_{11}$ phase of Tetiel \cite{Teitel1948}, but accurate density measurements exclude the possibility of a AB$_{11}$ composition.
Aldinger \etal \cite{Webster1979} used the structure of Johnson \etal \cite{Johnson1970} but assigned the composition FeBe$_7$, yet the compound is still presented with a composition of 8 at.~\% Fe. %Notably, this structure contains 19 atoms per unit cell.
Later, J\"onsson \etal \cite{Jonsson1982a} were able to index the same X-ray peaks to another hexagonal structure, with c/a ratio 1.50, rather then the previously reported value of 2.59.

The ternary Al-Be-Fe system is even less well characterised. Raynor \etal \cite{Raynor1953} report a phase with composition Fe$_3$Al$_7$Be$_7$. Black \cite{Black1955} presents an intermetallic with composition FeAl$_2$Be$_{2.3}$, which exhibits a defective form of the cubic C15 Laves structure (prototype MgCu$_2$) where Fe and Be atoms are ordered on the Cu site (the Be deficiency was not explained). Both studies concentrated on the Al-rich side of the phase diagram and therefore did not identify any low Al, high Be phases.

Subsequent work focused on commercial Be-rich alloys: Rooksby and Green \cite{Rooksby1962a} found an intermetallic similar to FeBe$_5$ but with a larger lattice parameter, which was initially termed YBe$_5$ (where Y indicates a transition metal, not yttrium), and was later identified as (Al,Fe)Be$_5$\cite{Rooksby1962}, in which the presence of Al on Fe sites results in a larger lattice parameter.
Later Carrabine \cite{Carrabine1963} presented a cubic AlFeBe$_4$ phase and in an addendum explained how this composition clarifies the results from Rooksby \cite{Rooksby1962}. It is not clear, however, whether the phase is ordered or disordered.
Myers and Smugeresky \cite{Myers1978} measured the maximum and minimum $\Al/\Fe$ atomic ratios of AlFeBe$_4$ ($1.4\pm0.1$ and $0.98\pm0.15$ respectively) and noted that the Be-rich $\varepsilon$ phase does not accommodate an appreciable amount of Al and that the stability of this phase reduces with increasing Al content.

The current work focuses on those phases that are relevant to Be alloys with low Al and Fe concentrations. Using density functional theory (DFT), we will consider the stability of Fe and Al as the FeBe$_2$, FeBe$_5$, $\varepsilon$ and AlFeBe$_4$ intermetallics, deviations from these stoichiometric compositions and the solubility of Fe and Al within Be metal.
The article is structured as follows: after describing the computational methodology, we provide an overview of the crystal structures that are relevant to the work, highlighting the similarities between the phases. 
The results are then presented in three main sections: first we examine the binary Fe-Be intermetallic compounds and by considering their relative stability, we predict the structure of the $\varepsilon$ phase to be hexagonal Fe$_2$Be$_{17}$. In subsections we consider elastic properties, magnetic contributions, temperature effects and the accommodation of non-stoichiometry. 
Next, we consider the effect of Al additions and the consequent formation of solid solutions and a ternary Al-Fe-Be phase. Again we consider magnetic properties and non-stoichiometry.
Next, the driving force for ordering in each of the intermetallic phases is presented, after which, we summarise our findings.

%%%%%%%%%%%%%       DFT     METHODOLOGY             %%%%%%%%%%%%
\section{Computational Methodology}
\label{sec:meth}
The DFT simulations employed the Perdew Burke and Ernzerhof (PBE) \cite{Perdew1996} formulation of the generalised gradient approximation for the exchange-correlation functional. Ultra-soft pseudo potentials with a consistent cut-off of \SI{400}{eV} were used throughout. All simulations were carried out using the {\sc castep} code \cite{Clark2005}.

For point defect calculations, a supercell consisting of $2 \times 2 \times 2$ conventional unit cells (containing 192 atoms) was used for the cubic AlFeBe$_4$, FeBe$_5$ and FeBe$_2$ phases, while a $3 \times 3 \times 2$ supercell (216 atoms) was employed for the C14 Laves hexagonal polymorph of FeBe$_2$ and a $3 \times 3 \times 1$ supercell for Fe$_2$Be$_{17}$ (171 atoms).
A high density of {\bf k}-points was used for the integration of the Brillouin Zone, following the Monkhost-Pack sampling scheme \cite{Monkhorst1976}: the distance between sampling points was maintained as close as possible to \SI{0.30}{nm^{-1}} and never above \SI{0.35}{nm^{-1}}. In practice this means a sampling grid of $3\times3\times3$ points for the largest supercells. % 0.03 Ang^-1 = 0.3 nm^-1 (reciprocal units)

Since these systems are metallic, density mixing and Methfessel-Paxton \cite{Methfessel1989} cold smearing of bands were employed with a width of \SI{0.1}{eV}. Testing was carried out to ensure a convergence of \SI{e-3}{eV/atom} with respect to all parameters.
No symmetry operations were enforced when calculating point defects and all calculations were spin polarised, taking particular care that defective cells reached the lowest energy magnetic state (see \ref{sec:mag} for further details).

The temperature dependence of thermodynamical quantities was calculated within the harmonic and quasi-harmonic approximations.
%The latter is an extension of the former, in which lattice expansion of the material is taken into account by calculating the phonon DOS at multiple unit cell volumes. Then, for each temperature the Birch-Murnaghan equation of state was fitted
In the harmonic approximation, the vibrational enthalpy $H_{vib}(T,V)$ --- which includes the zero-point energy (ZPE) --- and the vibrational entropy $S_{vib}(T,V)$ are evaluated by integrating the phonon DOS. Together with the configurational entropy $S_{conf}(T)$ and internal energy of the system $U(V)$, they provide Helmholtz free energy, $F(T,V)$:
\begin{align}
F(T,V) &= U(V) + F_{phonon}(T,V) - TS_{conf}(T) \\
F(T,V) &= U(V) + H_{vib}(T,V) - TS_{vib}(T,V) - TS_{conf}
\end{align}
where $V$ is volume and $T$ is temperature. $S_{conf}$ is computed using Boltzmann statistics:
\begin{align}
S_{conf} &= k_B \ln(\Omega)
\end{align}
where $\Omega$ is the number of possible states.

\begin{figure*}[hbt]
\centering
\includegraphics[width=0.8\textwidth]{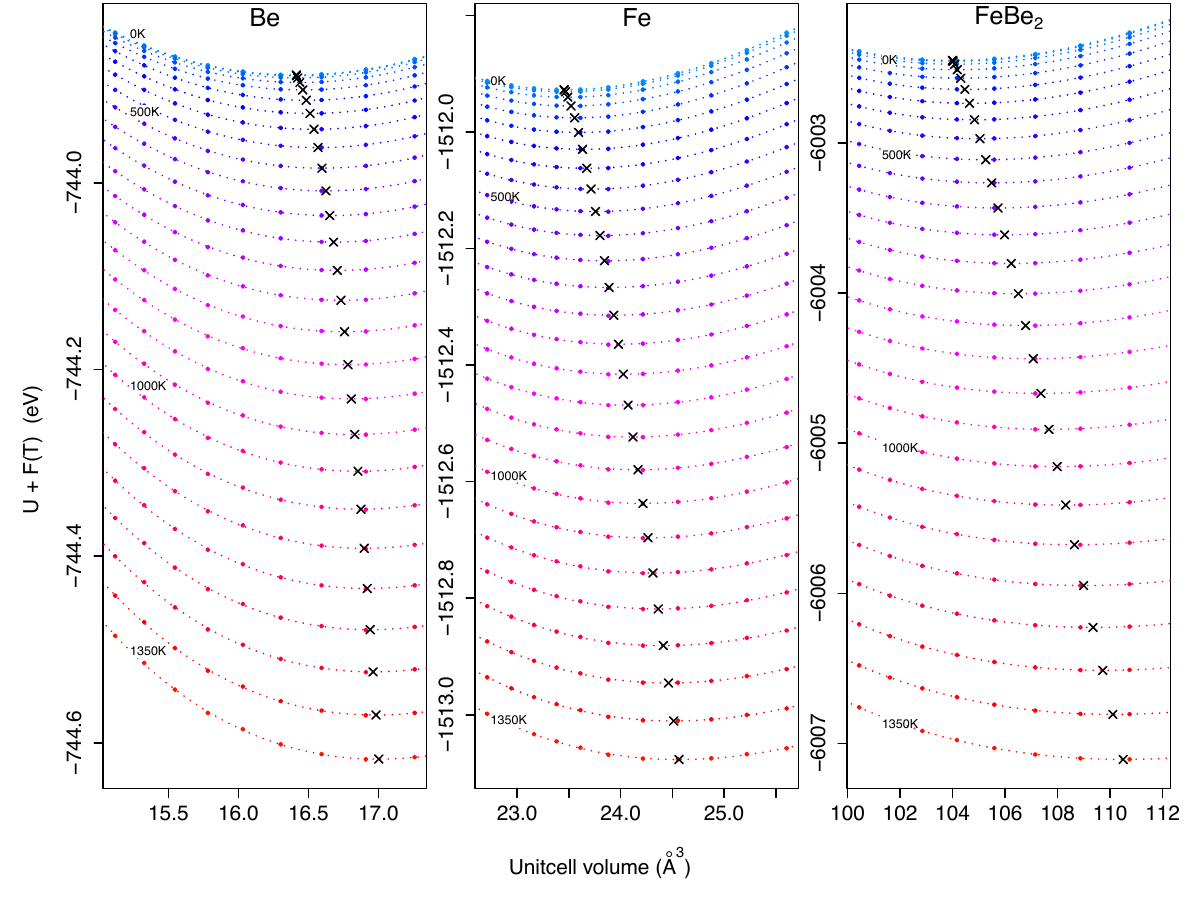}
\caption{\label{fig:quasiharm} Free energy vs volume. Crosses are represent the minimum energy volume for each temperature.}
\end{figure*}

Helmholtz free energy is relevant for constant volume conditions. However, more commonly, experiments are carried out at constant pressure conditions, in which Gibbs free energy is the relevant quantity.
The later may be obtained using the quasi-harmonic method, in which phonon DOS integration is performed at multiple unit cell volumes. Then, for each temperature, the Birch-Murnaghan \cite{Murnaghan1944,Birch1947b} equation of state is fitted to the free energy of the system and only the minimum energy value is taken (see Fig.~\ref{fig:quasiharm}):
\begin{equation}
G_{phonon}(T,P) = \min\limits_{V}\left( U(V) + F_{phonon}(T,V) \right)
\end{equation}

\begin{figure}[hbt]
\centering
\includegraphics[width=0.33\textwidth]{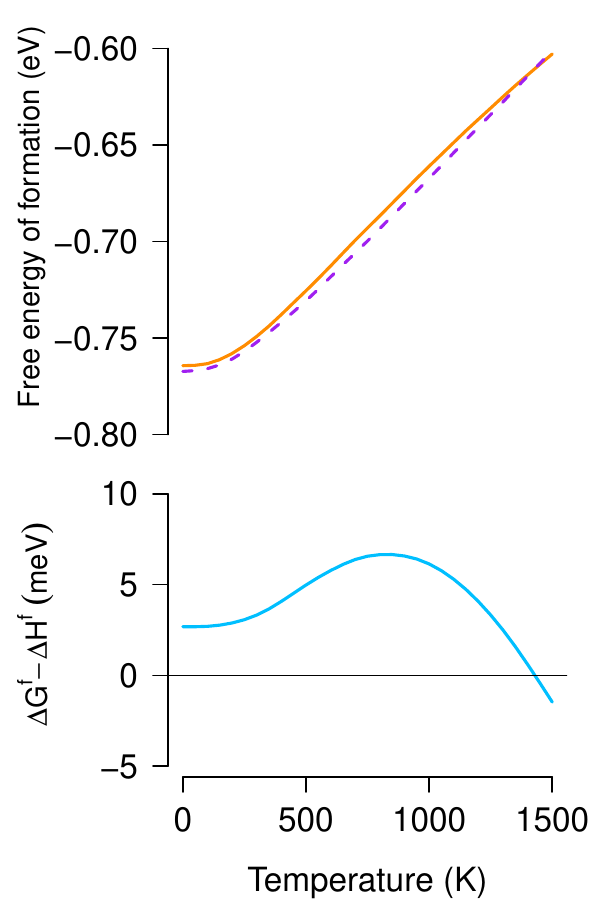}
\caption{\label{fig:GibHelzDiff} Free energy of formation calculated using the quasi-harmonic approximation ($\Delta G^f$, orange solid line) and the harmonic approximation ($\Delta H^f$, purple dashed line), both in eV. Bellow, in light blue is the difference ($\Delta G^f-\Delta H^f$), reported in meV.}
\end{figure}

Due to computational restrictions, the quasi-harmonic approximation was employed for the smallest systems under consideration, i.e.\ Fe(s), Be(s) and h-FeBe$_2$(s).
The use of the quasi-harmonic method yields elastic properties that are closer to the experimental values (see Table~\ref{tab:lattice}). However, the relative stability of the intermetallic was barely affected by the choice of method, as shown in Figure~\ref{fig:GibHelzDiff} where the scale is in meV; this provides confidence on the use of the harmonic approximation for the remaining phases.

Recent advances in the description of magnetic ordering \cite{Kormann2014,Lavrentiev2014,Kormann2010,Kormann2008,Lavrentiev2013,Ma2013} allow for the accurate calculation of magnetic contributions to the total stability of phases. This is particularly significant for iron. Since the current work focuses on the relative stability of intermetallic phases of the Be-Al-Fe system, accurately describing the absolute energy of the reference Fe phase is of less importance, as this is kept consistent across the study. Nevertheless, the use of advanced methods for magnetic materials may also be relevant to the intermetallic phases. To investigate that, a thorough analysis of the magnetic ordering of the phases presented here is require, which is beyond the scope of the current work.

%By replicating the same procedure at multiple unit cell volumes, then the Birch-Murnaghan equation of state may be used to predict the lowest energy volume for each temperature. Following that, Gibbs free energy may be computed by evaluating 
%In this work, $F(T,V)$ was calculated for the ground state volume of each cell, no anharmonic contributions were considered.
%Note that the current approach does not include the configurational entropy due to large variation in composition, therefore is limited to the stoichiometric or near-stoichiometric compositions (i.e.\ single point defects only).

\begin{table*}[hbt]
\centering
\small
\caption{ \label{tab:lattice} Simulated and experimental lattice parameters and bulk moduli.}
\begin{tabular}{l l S S S S S S}
\toprule
		&					&$a^{0K}$~\si{(\angstrom)}	&$a^{300K}$~\si{(\angstrom)}	&$c^{0K}$~\si{(\angstrom)}	&$c^{300K}$~\si{(\angstrom)}	&$K^{0K}$~\si{(GPa)}	&$K^{300K}$~\si{(GPa)}	\\
\midrule
Fe(s)		&ground state		&2.859	&\text{---}	&\text{---}	&\text{---}	&105\pm5		 &\text{---}	\\
		&quasi-harmonic	&2.863	&2.872	&\text{---}	&\text{---}	&195\pm2		&186\pm2	\\
		&\cite{Owen2002,Adams2006}	&2.8550	&2.8598	&\text{---}	&\text{---}	&170.4		&166.2	\\
\midrule
Be(s)	&ground state		&2.273	&\text{---}	&3.583	&\text{---}	&125\pm1	&\text{---}	\\
		&quasi-harmonic	&2.290	&2.293	&3.608	&3.613	&133\pm2 	&135\pm2	\\
		&\cite{Owen1952a}\cite{Smith1960}& NA&2.286&	NA	&3.585	&133.6	&131.2	\\
\midrule
FeBe$2$	&ground state		&4.179	&\text{---}	&6.799	&\text{---}	&161.5\pm1	&\text{---}	\\
		&quasi-harmonic	&4.195	&4.209	&6.820	&6.843	&153\pm2		&145\pm2	\\
		&\cite{Okamoto1988}&NA		&4.219\pm0.003&NA&6.856\pm0.008&NA	&NA		\\
%\midrule
%FeBe$5$	&ground state		&5.911	&\text{---}	&\text{---}	&\text{---}	&141.3	&\text{---}	\\
%		&quasi-harmonic	&		&		&\text{---}	&\text{---}	&		\\
%		&[ref] (RT)			&5.881\pm0.013&	&\text{---}	& \text{---}	&		\\
\bottomrule
\end{tabular}
\end{table*}

%\begin{figure}
%\centering
%\includegraphics[width=0.33\textwidth]{thermo_GibbsHelmh.pdf}
%\caption{\label{fig:GibHelzDiff} Difference in free energy of formation between the harmonic approximation (solid line) and quasi-harmonic approximation (dashed line).}
%\end{figure}

Phonon densities of states (DOS) were calculated using the finite displacement method with supercell extrapolation \cite{Frank1995}. Supercells containing 48, 162, 192 and 384 atoms were used to test convergence with respect to supercell size for FeBe$_2$. The difference in harmonic thermodynamical contribution between the 384 atom supercell and the 48 atom supercell was smaller than \SI{E-2}{eV/formula unit}. 

%\begin{table}
%\centering
%\small
%\color{red}
%\caption{ \label{tab:lattice} Simulated and experimental lattice parameters and bulk moduli}
%\begin{tabular}{l l S S S}
%\toprule
%		&					&$a$~\si{(\angstrom)}	&$c$~\si{(\angstrom)}	&$K$~\si{(GPa)}	\\
%\midrule
%Fe(s)		&ground state			&2.859			&\text{---}			&		\\
%		&quasi-harmonic (0K)	&2.863			&\text{---}			&194.9	\\
%		&quasi-harmonic (273K)	&2.872			&\text{---}			&184.2	\\
%		&[ref]				&				&\text{---}			&		\\
%\midrule
%Be(s)	&ground state			&2.273			&3.583			&		\\
%		&quasi-harmonic (0K)	&2.290			&3.608			&132.3	\\
%		&quasi-harmonic (273K)	&2.293			&3.613			&138.7	\\
%		&[ref]				&				&				&		\\
%\midrule
%FeBe$2$	&ground state			&4.179			&6.799			&161.5	\\
%		&quasi-harmonic (0K)	&4.195			&6.820			&152.8	\\
%		&quasi-harmonic (273K)	&4.209			&6.843			&144.1	\\
%		&[ref] (RT)				&4.219\pm0.003	&6.856\pm0.008	&		\\
%\midrule
%FeBe$5$	&ground state			&5.911			&\text{---}			&141.3	\\
%		&quasi-harmonic (0K)	&				&\text{---}			&		\\
%		&quasi-harmonic (273K)	&				&\text{---}			&		\\
%		&[ref] (RT)				&5.881\pm0.013	&\text{---}			&		\\
%\bottomrule
%\end{tabular}
%\end{table}

For defective supercells, the energy convergence criterion for self-consistent calculations was set to \SI{1e-8}{\electronvolt}.
Similarly robust criteria were imposed for atomic relaxation: the energy difference was less than \SI{1e-6}{\electronvolt}, forces on individual atoms less than \SI{ 0.01}{\electronvolt\per\angstrom} and for constant pressure calculations, the stress component on cells less than \SI{0.05}{\giga\pascal}.
For phonon calculations and ideal structures, the degree of convergence was tightened by 1.5 orders of magnitude.

Elastic constants were calculated using tools developed by Walker and Wilson~\cite{Walker} by performing small lattice perturbations from the ground state structures and measuring the stresses. Ten strain increments were performed in each crystallographic independent direction, between $-0.01$ and $0.01$. 
Theoretical XRD patterns were produced with CrystalDiffract$^{\circledR}$~\cite{CrystalMaker}, with a peak broadening of \SI{0.001}{\angstrom^{-1}}.

Employing the methodology developed by Bragg and Williams \cite{Bragg1934,Bragg1935,Williams1935a}, it is possible to estimate the degree of order ($S$) of a phase as a function of temperature and potential energy increase ($V$) caused by an atomic replacement from order towards disorder.
The degree of order of a structure may be defined as follows: let $N$ be the total number of atoms in the system, and $n$ the subset of atoms that are susceptible to disordered substitutions. Further, let there be $rn$ positions of order in the system and therefore $(1-r)n$ positions of disorder and let $p$ be the probability that an atom is occupying a position of disorder. The degree of order $S$ is then defined as:
\begin{align}
S=&
\frac{\text{\tiny{actual value of $p$ $-$ value of $p$ for complete disorder}}}{\text{\tiny{value of $p$ for complete order $-$ value of $p$ for complete disorder}}}\nonumber	\\
=&	\frac{p-r}{1-r}
\end{align}
so that in complete disorder (i.e.\ when $p=r$) $S=0$, and in complete order (i.e.\ when $p=1$) $S=1$.

The systems considered in the current work exhibit the same order/disorder parameters of Fe$_3$Al examined in ref.~\cite{Bragg1934}, namely $n=\sfrac{N}{2}$, $r=\sfrac{1}{2}$ and the total number of A atoms (Al in the current work) $= rn$.
With this set of parameters, the dependency of the degree of order $S$ with temperature $T$ and potential energy $V$ of a replacement towards disorder, can be simplified to~\cite{Bragg1934}:
\begin{align}
S(V,T) = \tanh(x/4),&&x=V(S,T)/k_BT
\end{align}
where $k_B$ is Boltzmann constant.
The energy penalty $V$ is in turn dependant on the degree of order $S$. In the Bragg-Williams approach this dependency is assumed to be linear. Furthermore if $S=0$ (complete disorder), $V$ must also be zero as the positions of order and those of disorder are indistinguishable and substitutions into either site must be equivalent. Owing to the linear relationship, $V$ reaches a maximum value $V_0$ when $S=0$ (i.e.\ in conditions of complete order). Mathematically, that is expressed as $$V(S,T)=V_{0}S(V,T),$$ so that $$V(0,T)=0$$ and $$V(1,T)=V_{0}.$$
It is acknowledged that local fluctuation of the atomic arrangements in any small sample of crystal will cause corresponding fluctuation in $V$, therefore $V$ is to be taken as an \emph{effective average} value of V, representative of the degree of order S \cite{Williams1935a}.
Bragg and Williams also recognised that $V$ is almost insensitive to $T$ \cite{Bragg1934}. In the current work the temperature dependency of $V$ is ignored altogether and $V_{0}$ is taken (for all temperatures) as half the average antisite defect formation energy in a completely ordered crystal.

Similarly to the calculation of magnetic disorder described above, other methods for the computation of chemical disorder transition have been proposed subsequent to the Bragg-Williams approach adopted here \cite{Kikuchi1951,Kikuchi1985,VandeWalle2002,VandeWalle2002a,Kormann2014},  however these method require a great deal more computational power and separate in-depth analysis, and are outside the scope of this study.

 %%%%%%%%%%%%%%%      XTALLOGTAPHY      %%%%%%%%%%%%%%%
\section{Crystallography of Fe-Al-Be intermetallics}
\label{sec:xtal}

\begin{figure}[htb]
\centering
%\begin{subfigure}[b]{0.49\columnwidth}
\begin{subfigure}[b]{0.245\columnwidth}
                \centering
                \includegraphics[width=\textwidth]{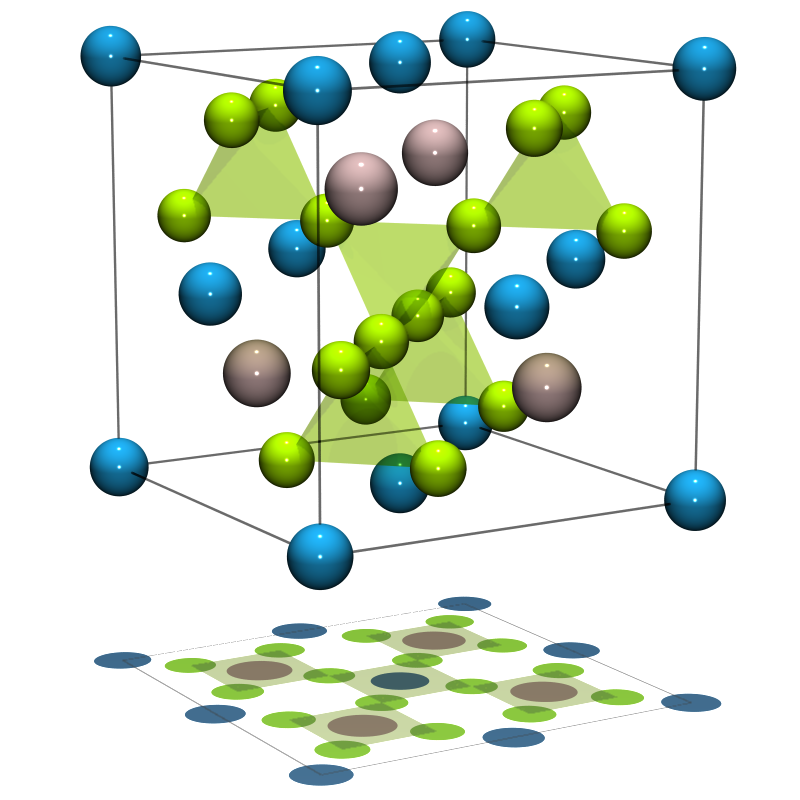}
                \caption{AlFeBe$_4$}
                \label{fig:AlFeBe4}
\end{subfigure}
%\begin{subfigure}[b]{0.49\columnwidth}
\begin{subfigure}[b]{0.245\columnwidth}
                \centering
                \includegraphics[width=\textwidth]{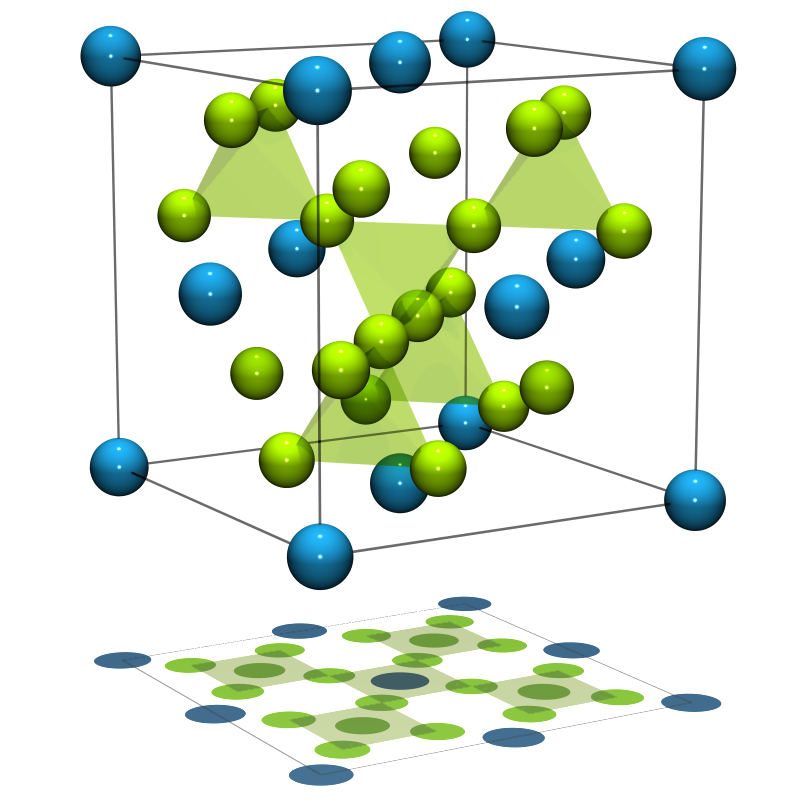}
                \caption{FeBe$_5$}
                \label{fig:FeBe5}
\end{subfigure}
\\
%\begin{subfigure}[b]{0.49\columnwidth}
\begin{subfigure}[b]{0.245\columnwidth}
                \centering
                \includegraphics[width=\textwidth]{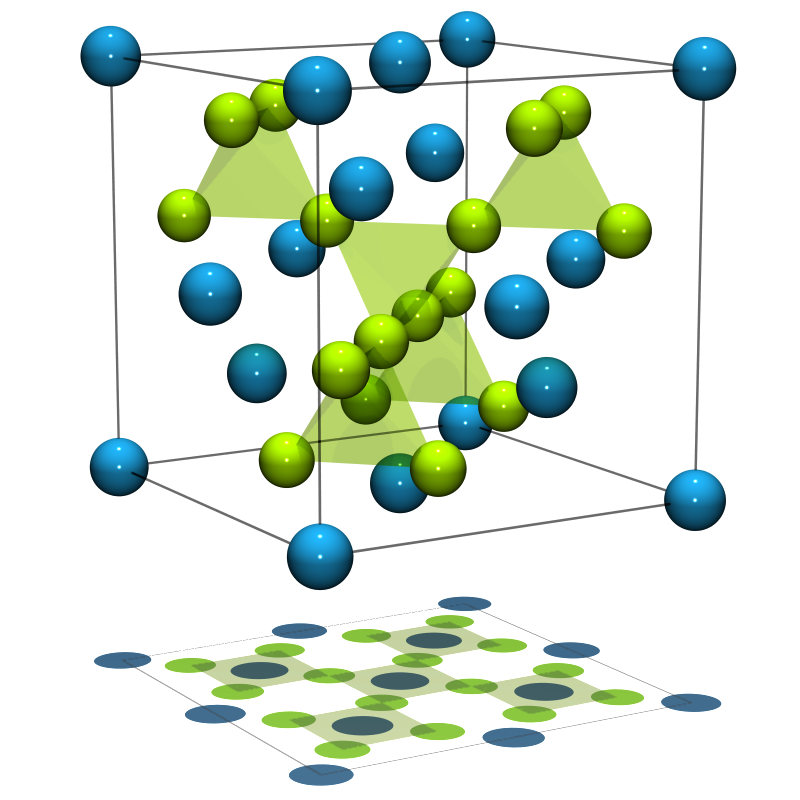}
                \caption{FeBe$_2$-C15}
                \label{fig:FB2_C15}
\end{subfigure}
%\begin{subfigure}[b]{0.49\columnwidth}
\begin{subfigure}[b]{0.245\columnwidth}
                \centering
                \includegraphics[width=\textwidth]{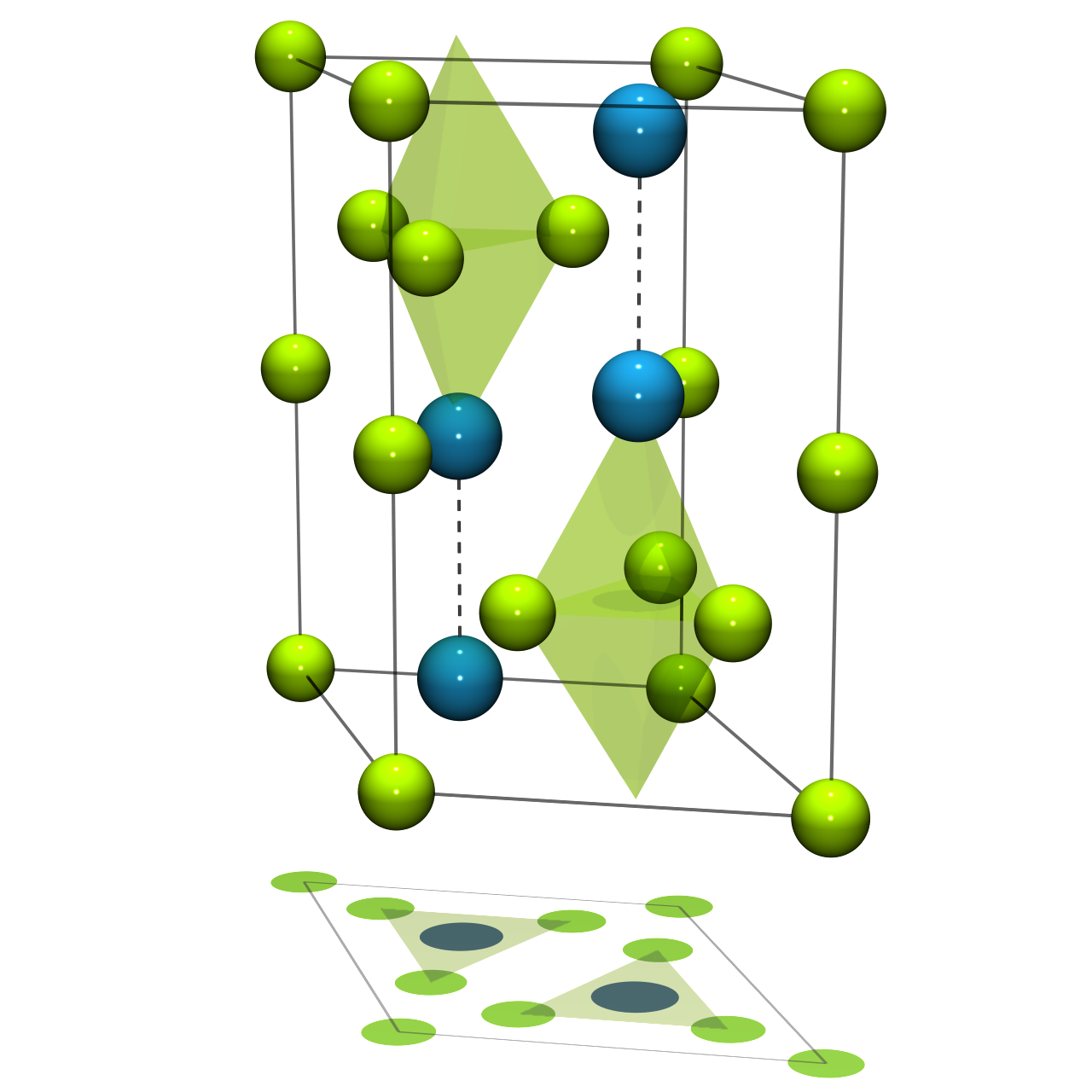}
                \caption{FeBe$_2$-C14}
                \label{fig:FB2_C14}
\end{subfigure}
\caption{\label{fig:structures} Crystal structures of (a) AlFeBe$_4$, (b) FeBe$_5$, (c) the cubic C15 Laves phase of FeBe$_2$, with prototype Cu$_2$Mg structure and (d) the hexagonal C14 structure of FeBe$_2$, with prototype MgZn$_2$ structure. Smaller green atoms are Be, the larger blue atoms are Fe and the larger pale pink atoms are Al.}
\end{figure}

The AlFeBe$_4$ phase (space group $F\bar{4}3m$), can be described by three face-centered cubic (FCC) sublattices (see Fig~\ref{fig:AlFeBe4}). The first sublattice, with origin at $(0, 0, 0)$ is occupied by Fe atoms. The second one, occupied by Al atoms, is shifted by $\left[\tfrac{3}{4} \tfrac{1}{4} \tfrac{1}{4}\right]$ and has four sites within the conventional unit cell, namely $\left(\tfrac{3}{4}, \tfrac{1}{4}, \tfrac{1}{4}\right), \left(\tfrac{1}{4}, \tfrac{3}{4}, \tfrac{1}{4}\right), \left(\tfrac{1}{4}, \tfrac{1}{4}, \tfrac{3}{4}\right) \text{ and } \left(\tfrac{3}{4}, \tfrac{3}{4}, \tfrac{3}{4}\right)$.
Be atoms, which have a multiplicity of 4 compared to Al or Fe atoms, are grouped in tetrahedra, each of which is centred at the lattice points of the third FCC sublattice.
The third FCC sublattice is shifted by $\left[\tfrac{1}{4} \tfrac{1}{4} \tfrac{1}{4}\right]$, thereby occupying the last four corners of the cube: $\left(\tfrac{1}{4}, \tfrac{1}{4}, \tfrac{1}{4}\right), \left(\tfrac{3}{4}, \tfrac{3}{4}, \tfrac{1}{4}\right), \left(\tfrac{1}{4}, \tfrac{3}{4}, \tfrac{3}{4}\right) \text{ and } \left(\tfrac{3}{4}, \tfrac{1}{4}, \tfrac{3}{4}\right)$.

FeBe$_5$ exhibits the same structure as AlFeBe$_4$, where all Al atoms have been substituted for Be (Fig~\ref{fig:FeBe5}).
If all Al atoms were to be substituted by Fe instead, the structure would become the C15 Laves phase of FeBe$_2$ (Fig.~\ref{fig:FB2_C15}).
Experimentally it has been reported that FeBe$_2$ exhibits the C14 Laves phase (Fig.~\ref{fig:FB2_C14}). Nevertheless, as a check of the validity of the current methodology, the C15 structure was also modelled. Although the two polymorphs of FeBe$_2$ may look very different, the local atomic coordination is the same: the A atoms (either Fe or Al in the current work) form a diamond structure sub-lattice, where each atom had a coordination number (CN) of 16 (4 A atoms and 12 B atoms). The B atoms (Be) form a network of tetrahedra that intercalate around the A atoms, with a CN of 12 (6 A + 6 B).

A disordered Al baring phase has also been reported where Al substitutes for Fe in FeBe$_5$, producing (Al,Fe)Be$_5$ \cite{Rooksby1962}. In the current work we also consider the case in which Al substitutes for Fe in FeBe$_2$, forming (Al,Fe)Be$_2$. In both cases, the ternary compounds retain the lattice symmetry of their parent structures.

\begin{alphafootnotes}
\begin{table*}[hbt]
\small
\centering
\caption{\label{tab:ref_epsilon} Summary of crystallographic information available regarding the Be-rich $\varepsilon$ phase.}
\begin{tabular}{l l l l c S S r}
\toprule
\multirow{2}{*}{Composition}	&Crystal	&Prototype&Space	&Atoms per	&$a$&	$c$	&\multirow{2}{*}{Reference}	\\
						&class	&structure&group	&unit cell		&\si{(\angstrom)}&\si{(\angstrom)}			\\ \midrule
FeBe$_9$					&---		&---		&---		&---			&\text{---}	&\text{---}	&\cite{Teitel1948}\protect\footnotemark[1]\\
FeBe$_{11}$				&Hex	&---		&---		&18			&4.13	&10.71	&\cite{Teitel1948}	\\
FeBe$_{12}$				&Tetr		&Mn$_{12}$Th	&$I_4mmm$	&13	&4.323	&7.25	&\cite{VonBatchelder1957}\\
FeBe$_{11}$				&Hex	&---		&---		&---			&4.13	&10.72	&\cite{Rooksby1962}	\\
FeBe$_x$					&Hex	&RhBe$_{6.6}$	&$P\bar{6}m2$	&19\protect\footnotemark[2]	&4.137	&10.72	&\cite{Johnson1970}\\
FeBe$_{11}$				&Hex	&---		&---		&---			&4.13	&10.72	& \cite{Levine1971}	\\
FeBe$_7$					&Hex	&---		&---		&---			&7.13	&10.99	&\cite{Webster1979}	\\
FeBe$_{11}$				&Hex	&---		&---		&---			&7.15	&10.72	&\cite{Jonsson1982a}	\\
\bottomrule
\end{tabular}
\end{table*}

\footnotetext[1]{The work of Mish was not published but is indirectly reported in \cite{Teitel1948}.}
\footnotetext[2]{While the phase by Johnson \etal \cite{Johnson1970} has 19 symmetry sites, not all are fully occupied.}

\begin{table*}[hbt]
\small
\centering
\caption{\label{tab:epsilon} Crystal structures modelled to replicate the $\varepsilon$ phase. Formation enthalpy $H_f$ are normalised per Fe atom.}\begin{tabular}{l l l l c S S S S}
\toprule
\multirow{2}{*}{Composition}	&Crystal	&Prototype		&Space		&Atoms per	&$a$	&$c$		&$H_f$	\\%&$\Delta H_f$\\
						&class	&structure			&group		&unit cell		&\si{(\angstrom)}&\si{(\angstrom)}&\si{(eV)}\\%&\si{(eV)}\\ 
\midrule
Fe$_3$Be$_{16}$	&Hex	&RhBe$_{6.6}$				&$P\bar{6}m2$	&19	&4.08	&10.73	&-0.61	\\%&0.36\\
FeBe$_8$	~(1)		&Hex	&RhBe$_{6.6}$				&$P\bar{6}m2$	&18	&4.09	&10.72	&-0.13	\\%&0.84\\
FeBe$_8$	~(2)		&Hex	&RhBe$_{6.6}$				&$P\bar{6}m2$	&18	&4.10	&10.63	&-0.20	\\%&0.77\\
$\boldsymbol{\text{\bf{Fe}}_2 \text{\bf{Be}}_{17} }$	&\bf Hex	&{\bf RhBe}$_{\boldsymbol{6.6}}$	&$\boldsymbol{P\bar{6}m2}$	&$\boldsymbol{19}$	&$\boldsymbol{4.10}$	&$\boldsymbol{10.63}$	&$\boldsymbol{-0.97}$	\\%&	\text{\bf{---}}\\
Fe$_2$Be$_{15}$	&Hex	&RhBe$_{6.6}$				&$P\bar{6}m2$	&17	&4.15	&10.45	&0.71	\\%&1.68\\
FeBe$_{17}$		&Hex	&RhBe$_{6.6}$				&$P3m1$		&18	&4.11	&10.64	&1.20	\\%&2.17\\
FeBe$_{12}$		&Tetr		&Mn$_{12}$Th				&$I_4mmm$	&13	&7.16	&4.09	&-0.30	\\%&0.67\\
FeBe$_{12}$		&Hex	&Fe$_6$Ge$_6$Mg			&$P6/mmm$	&13	&4.15	&7.16	&0.56	\\%&1.53\\
FeBe$_{13}$		&Cubic	&NaZn$_{13}$				&$Fm\bar{3}c$	&28	&6.98	&\text{---}	&2.66	\\%&3.63\\
Fe$_2$Be$_{17}$	&Hex	&Ni$_{17}$Th$_2$			&$P6_3/mmc$	&38	&7.11	&7.04	&-0.15	\\%&0.82\\
Fe$_2$Be$_{17}$	&Hex	&Th$_2$Zn$_{17}$			&$R\bar{3}m$	&57	&5.41	&\text{---}	&-0.20	\\%&0.78\\
Fe$_3$Be$_{17}$	&Cubic	&Be$_{17}$Ru$_3$			&$Im\bar{3}$	&160	&10.99	&\text{---}	&-0.82	\\%&0.14\\
Be$_{22}$Fe		&Cubic	&Al$_{18}$Cr$_2$Mg$_3$	&$Fd\bar{3}m$	&176	&11.43	&\text{---}	&0.10	\\%&1.07\\
\bottomrule
\end{tabular}
\end{table*}

Regarding the Be-rich $\varepsilon$ phase, the limited crystallographic information available is summarised in Table~\ref{tab:ref_epsilon}.
In terms of the basis, Von Batchelder and Raeuchle \cite{VonBatchelder1957} provide a full set of atomic coordinates for the tetragonal FeBe$_{12}$ structure, but the only information available about the more commonly observed FeBe$_{11}$ phase of Teitel and Cohen \cite{Teitel1948}, is that a unit cell contains ${\sim}18$ atoms. 
The structure reported by Johnson \etal \cite{Johnson1970}, in a publication that focussed on the structure of RhBe$_{6.6}$, comprises a list of 9 atomic coordinates which, if fully occupied, would yield composition Fe$_3$Be$_{16}$. However, the exact composition of the compound (FeBe$_x$) was not provided, and some partial occupancy may be present on selected Fe and Be sites.
Interestingly, the phase described by Johnson \etal \cite{Johnson1970} shares similarities with that reported by Teitel and Cohen \cite{Teitel1948} for FeBe$_{11}$; this would explain the presence of ${\sim}18$ atoms per unit cell of FeBe$_{11}$. 
Aldinger \cite{Webster1979} and J\"onsson \etal \cite{Jonsson1982a} report structures with a larger lattice constant but do not give information regarding the crystal basis.

In the current work, we considered tetragonal FeBe$_{12}$, hexagonal Fe$_3$Be$_{16}$ and some variations of this structure that were generated by removing or changing those atoms that may accommodate partial occupancy (as observed in RhBe$_{6.6}$). Consequently two variants of FeBe$_8$, Fe$_2$Be$_{17}$, FeBe$_{17}$ and Fe$_2$Be$_{15}$ were modelled.
Furthermore, we have considered the structures of intermetallic compounds that Be can form with any other transition metal in which the ratio of transition-metals to Be is smaller than $\sfrac{1}{6}$. A summary of all the phases simulated, and their calculated enthalpies of formation from standard state $H_f$, are presented in Table~\ref{tab:epsilon}.
The reference state for Fe was the ferromagnetic body-centred cubic (ferrite), as per reference \cite{Herper1999}.
For small compositional variations such as those in Table~\ref{tab:epsilon}, the reaction energy to go from one phase to another phase under Be rich conditions
can be approximated by the change in formation enthalpy (to within \SI{0.005}{eV}).

The Fe$_2$Be$_{17}$ phase exhibits the lowest enthalpy of formation, with predicted lattice parameters in good agreement with previous work \cite{Teitel1948,Rooksby1962,Johnson1970,Levine1971}. All other variations of the RhBe$_{6.6}$ structures yielded less favourable formation energies and were not considered further.
None of the structures replicated from other transition metal beryllides proved to be more stable and were also discounted.
The tetragonal phase of Von Batchelder and Raeuchle \cite{VonBatchelder1957} is also significantly less favourable than hexagonal Fe$_2$Be$_{17}$. 

\begin{table}
\small
\centering
\caption{ \label{tab:epsilon_basis} Crystallographic basis parameters for $\varepsilon$-Fe$_{2-x}$Be$_{17+x}$.}
\begin{tabular}{l c c c c c}
\toprule
Atom		&Wyckoff		&$x$					&$y$					&$z$				&Occupancy	\\
label		&site			&					&					&				& \%			\\
\midrule
Fe1		&$2g$		&$\scriptstyle0$		&$\scriptstyle0$		&$\scriptstyle 0.192$	&76Fe + 24Be	\\
Be1		&$1d$		&$\sfrac{1}{3}$			&$\sfrac{2}{3}$			&$\sfrac{1}{2}$		&100Be	\\
Be2		&$1f$		&$\sfrac{2}{3}$			&$\sfrac{1}{3}$			&$\sfrac{1}{2}$		&100Be	\\
Be3		&$2g$		&$\scriptstyle0$		&$\scriptstyle0$		&$\scriptstyle0.400$	&100Be	\\
Be4		&$2h$		&$\sfrac{1}{3}$			&$\sfrac{2}{3}$			&$\scriptstyle0.131$	&100Be	\\
Be5		&$2i$		&$\sfrac{2}{3}$			&$\sfrac{1}{3}$			&$\scriptstyle0.156$	&100Be	\\
Be6		&$3j$		&$\scriptstyle0.8385$	&$\scriptstyle0.677$		&$\scriptstyle0$	&100Be	\\
Be7		&$6n$		&$\scriptstyle0.499$		&$\scriptstyle-0.499$		&$\scriptstyle0.3146$&100Be	\\
\bottomrule
\end{tabular}
\end{table}

The full crystallographic basis set for the Fe$_2$Be$_{17}$ structure is presented in Table~\ref{tab:epsilon_basis}. This includes partial occupancy of the Fe1 site, which is discussed in detail in section~\ref{sec:non-stoich1}.
Using the data presented in Tables \ref{tab:epsilon} and \ref{tab:epsilon_basis}, a theoretical XRD pattern was generated (green dashed line in Fig.~\ref{fig:epsilon_XRD}) and compared with the available experimental data from Rooksby \cite{Rooksby1962} (red solid line in Fig.~\ref{fig:epsilon_XRD}).
Localised models for exchange-correlation functionals (including the PBE used in the current work) are known to suffer from overbinding errors~\cite{Kohn1996,VandeWalle1999}, which in turn cause a shift in the XRD spectrum towards larger $\frac{1}{d}$ values. To compensate for this, a second spectrum (blue dotted line in Fig.~\ref{fig:epsilon_XRD}) was produced by enforcing the experimental lattice parameters of Rooksby \cite{Rooksby1962} with the predicted Fe$_2$Be$_{17}$ structure.
The excellent agreement between the experimental and theoretical XRD spectra further supports the conclusion that the defective Fe$_{2-x}$Be$_{17+x}$ structure is a good representation of the $\varepsilon$ phase (see section~\ref{sec:non-stoich1} for details on non-stoichiometry).

\begin{figure*}[htb]
\centering
	\includegraphics[width=\textwidth]{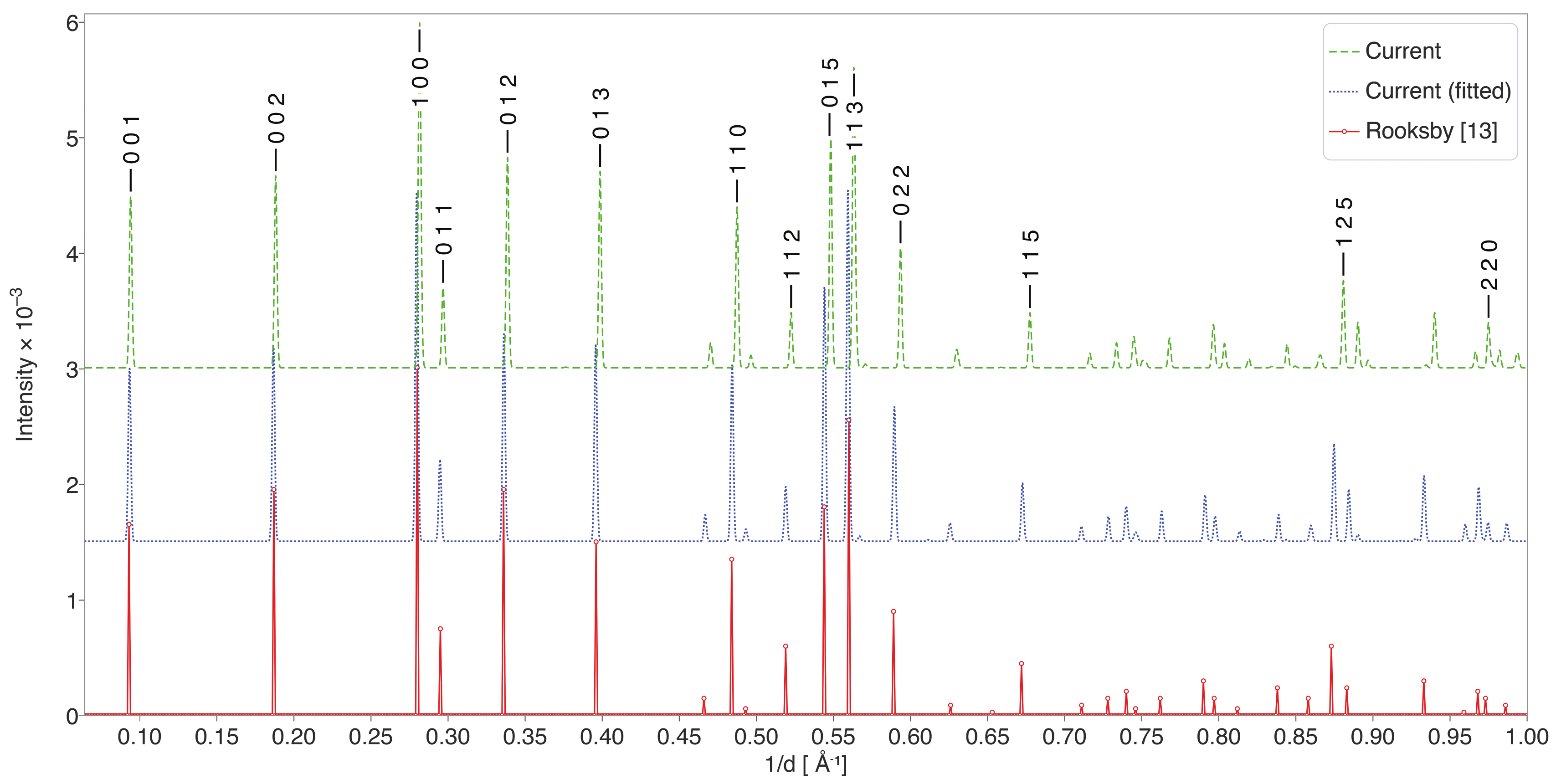}
	\caption{
	\label{fig:epsilon_XRD} Theoretical XRD spectra of Fe$_2$Be$_{17}$ and comparison with the observed spectra of the $\varepsilon$ phase, reproduced from the tabulated data of ref \cite{Rooksby1962}.}
\end{figure*}

%%%%%%%%%%%%           Result sections

\section{The binary Be-Fe system}
\label{sec:FeBe}
\subsection{Stability of the intermetallic phases}
\label{sec:binary_stability}
The enthalpy of formation from standard state of each phase under consideration was calculated following the generic reaction $\Fe + x \Be \rightarrow \FeBe_x$; these energies are presented alongside reactions~\ref{Ef(FeBe2)}--\ref{Ef(FBeps)}. For comparison, the solution enthalpy of Fe into Be metal is also presented (reaction~\ref{Esol(Fe_Be)}).
\begin{align}
\label{Ef(FeBe2)}
\Fe + 2 \Be 			&\xrightarrow{\SI{-0.81}{eV}}	 \FeBe_2\\
\label{Ef(FeBe5)}
\Fe + 5 \Be 			&\xrightarrow{\SI{-0.44}{eV}}	 \FeBe_5\\
\label{Ef(FeBe12)}
\Fe + 12 \Be 		&\xrightarrow{\SI{-0.30}{eV}}	 \FeBe_{12}\\
\label{Ef(FBeps)}
 \Fe + \tfrac{17}{2} \Be 		&\xrightarrow{\SI{-0.97}{eV}}	 \tfrac{1}{2} \Fe_{2}\Be_{17}\\
\label{Esol(Fe_Be)}
\Fe\s + \Be_{\Be} 		&\xrightarrow{\SI{-0.13}{eV}}	 \Fe_{\Be} + \Be\s 
\end{align}
Reactions are normalised per Fe atom.
For dilute Be solution, only the substitutional $\Fe_\Be$ species was considered (in a supercell containing 150 Be atoms), since previous work showed this to be the most favourable defect for the accommodation of Fe in Be \cite{Middleburgh2011}.

All phases exhibit favourable (negative) formation enthalpies, and in all cases these are lower than the enthalpy of solution. To better understand the relative stability of the intermetallics, the normalised formation enthalpies are plotted against composition to form a convex hull diagram (see Fig.~\ref{fig:convex}).
In such a diagram, the distance from the convex hull indicates the degree of instability of a phase, with the points lying on the hull identifying the phases that are observed at that composition \cite{Perevoshchikova2012,Lumley2014}.
Fig.~\ref{fig:convex} is constructed exclusively in terms of ground state enthalpy of each phase, with no considerations of entropic or temperature-dependant contributions, which will be presented later in section~\ref{sec:thermoFeBe}.

\begin{figure}[htb]
\centering
	\includegraphics[width=0.4\textwidth]{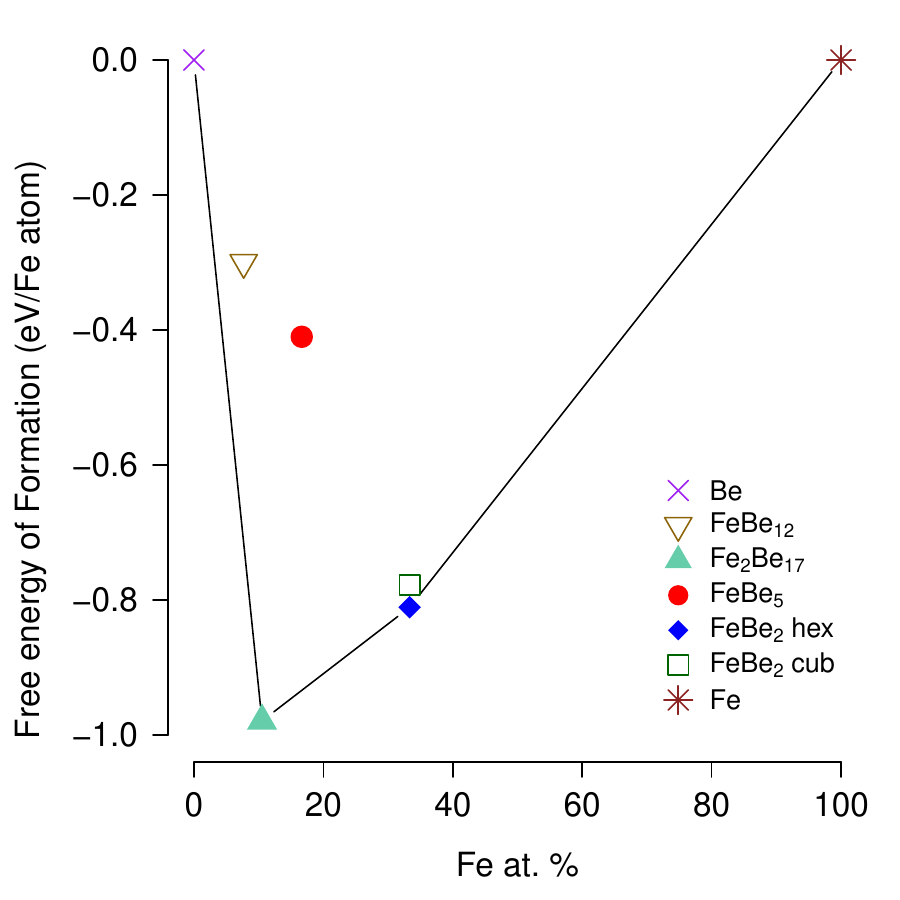}
	\caption{\label{fig:convex} Change in formation enthalpy with increasing Fe content. The line represent the convex hull and determines those phases that are expected to form.}
\end{figure}

If excess Fe is present, then FeBe$_2$ will be the predominant intermetallic phase observed in the alloy. This is supported by recent experimental observations by Kadyrzhanov \etal \cite{Kadyrzhanov2013}.
To quantify the driving force for the formation of FeBe$_2$, in the presence of excess Fe, reactions~\ref{Ef(FeBe2)}--\ref{Ef(FBeps)} can be rearranged to form reactions~\ref{FB5-FB2_ferrite}--\ref{FB12-FB2_ferrite}.
\begin{align}
	%\label{FB12-FBeps_ferrite}\FeBe_{12} + \Fe + 5 \Be	&\xrightarrow{\SI{-1.64}{eV}} 	\Fe_{2}\Be_{17}	\\
	%\label{FB12-FB5_ferrite}	\FeBe_{12} + \Fe	&\xrightarrow{\SI{-0.59}{eV}} 	2 \FeBe_5 + 2 \Be	\\
	\label{FB5-FB2_ferrite}	\FeBe_5 + \Fe	&\xrightarrow{\SI{-1.10}{eV}} 	2 \FeBe_2 + \Be	\\
	%\label{FBeps-FB5_ferrite}	\Fe_{2}\Be_{17} + \Fe	&\xrightarrow{\SI{+0.61}{eV}} 	3 \FeBe_5 + 2 \Be	\\
	\label{FBeps-FB2_ferrite}	\tfrac{1}{2} \Fe_{2}\Be_{17} + \Fe	&\xrightarrow{\SI{-0.24}{eV}} 	2 \FeBe_2 + \tfrac{9}{2} \Be	\\
	\label{FB12-FB2_ferrite}	\FeBe_{12} + \Fe	&\xrightarrow{\SI{-1.32}{eV}} 	2 \FeBe_2 + 8 \Be  
\end{align}
All reactions are exothermic (negative).
However, in the framework of Be alloys, the presence of excess Fe is unlikely, therefore reactions~\ref{FB5-FB2_Be}--\ref{FB12-FB2_Be} should also be considered, in which each phase ejects Be atoms (released into the Be bulk) to form a higher Fe content intermetallic phase.
\begin{align}
	%\label{FB12-FBeps_Be}\FeBe_{12}					&\xrightarrow{\SI{-0.67}{eV}}	\tfrac{1}{2} \Fe_2\Be_{17} + \tfrac{1}{2} \Be	\\
	%\label{FB12-FB5_Be}\FeBe_{12}					&\xrightarrow{\SI{-0.14}{eV}}	\FeBe_5 + 7 \Be	\\
	\label{FB5-FB2_Be}\FeBe_5						&\xrightarrow{\SI{-0.36}{eV}}	\FeBe_2 + 3 \Be	 \\
	%\label{FBeps-FB5_Be}\tfrac{1}{2} \Fe_{2}\Be_{17}		&\xrightarrow{\SI{+0.52}{eV}}	\FeBe_5 + \tfrac{7}{2} \Be	\\
	\label{FBeps-FB2_Be}\tfrac{1}{2} \Fe_{2}\Be_{17}		&\xrightarrow{\SI{+0.16}{eV}}	\FeBe_2 + \tfrac{13}{2} \Be	\\
	\label{FB12-FB2_Be}\FeBe_{12}					&\xrightarrow{\SI{-0.50}{eV}}	\FeBe_2 + 10 \Be 
\end{align}
Reaction~\ref{FBeps-FB2_Be}, shows that Fe$_2$Be$_{17}$ does not spontaneously decompose into FeBe$_2$ in dilute Fe-Be alloys. This is demonstrated in Fig.~\ref{fig:convex} because Fe$_2$Be$_{17}$ lies on the convex hull.
On the other hand, FeBe$_{12}$ and FeBe$_5$ phases appear to be well above the convex hull, suggesting that they are unstable at \SI{0}{K}.
With regards to FeBe$_{12}$, this is a strong indication that the the $\varepsilon$ phase exhibits the hexagonal (Fe$_2$Be$_{17}$) structure of Teitel and Cohen \cite{Teitel1948}, as reported in the majority of the literature, and not the tetragonal FeBe$_{12}$ structure suggested by von Batchelder and Raeuchle \cite{VonBatchelder1957}.

The instability of FeBe$_5$ is surprising considering the experimental observations of this phase \cite{Teitel1948,Okamoto1988,Rooksby1962,Webster1979}. However, this analysis is strictly related to the ground state enthalpy of the phases. Temperature effects and the presence of extrinsic point defects may stabilise FeBe$_5$, as addressed in sections~\ref{sec:thermoFeBe} and \ref{sec:diluteAl}.

Magnetic and elastic properties of all binary intermtallics were evaluated and the results are presented in \ref{sec:mag} and \ref{sec:elastics}, respectively. In particular, it was found that FeBe$_2$, FeBe$_5$ and Fe$_2$Be$_{17}$ are ferromagnetic (the former two particularly strongly).

\subsection{Temperature effects}
\label{sec:thermoFeBe}

Temperature dependent thermodynamic properties were calculated, within the harmonic approximation, by integrating the phonon DOS and subsequently including configurational entropy.
The Helmholtz free energy of formation $F_f(T,V)$ was calculated following reactions~\ref{Ef(FeBe2)}--\ref{Esol(Fe_Be)} and the results are presented for the temperature range of \SI{0}{K}--{1600}{K} (see Fig.~\ref{fig:thermo}).

\begin{figure}[htb]
\centering
	\includegraphics[width=0.5\textwidth]{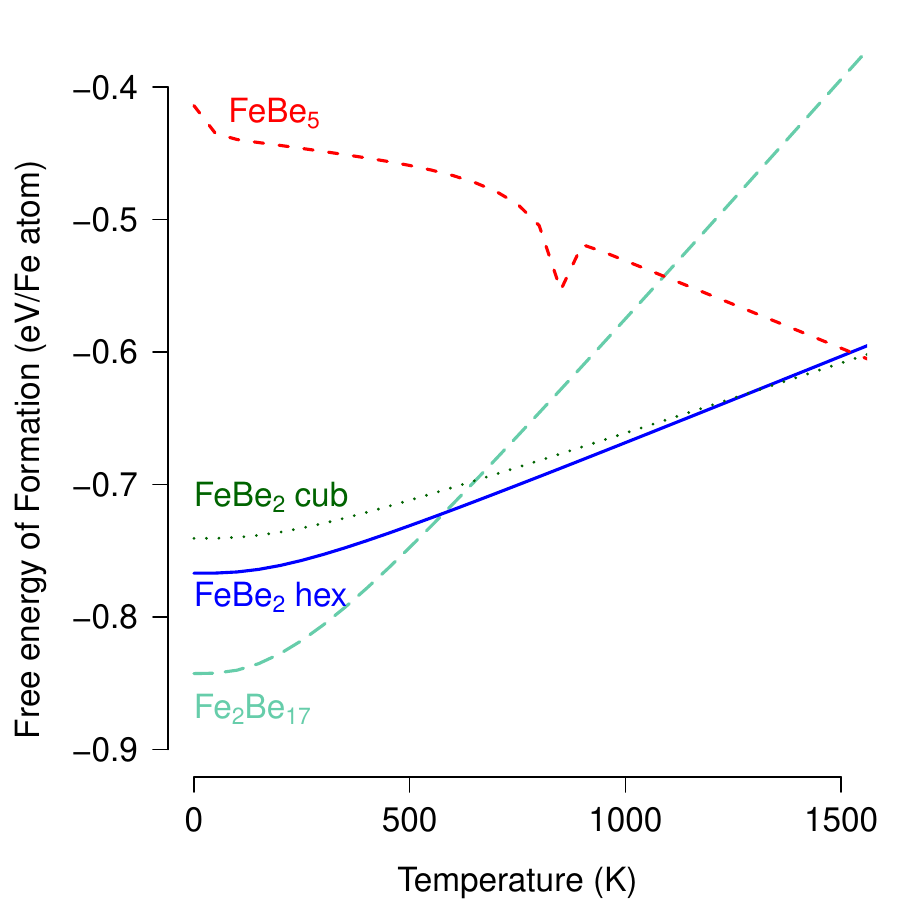}
	\caption{\label{fig:thermo} Free energy of formation versus temperature for Fe-Be binary intermetallics. The values are normalised per Fe atom.}
\end{figure}

The most striking feature of Fig.~\ref{fig:thermo} is that with increasing temperature, the stability of the FeBe$_5$ phase increases, while those of the other phases decrease. Thus, FeBe$_5$ is stabilised by temperature effects, although it is not expected to form at low temperature under equilibrium conditions. The spike observed in the free energy curve of FeBe$_5$ is due to an order/disorder transition, discussed in greater detail in section~\ref{sec:disorder}.

At ${\sim}1200$~K, the FeBe$_5$ curve crosses the Fe$_2$Be$_{17}$ curve. This is a condition necessary but not sufficient for the spontaneous decomposition of Fe$_2$Be$_{17}$ into FeBe$_5$, as the reaction energy also depends on the Fe content. Therefore, \SI{1200}{K} may be considered as the lower bound or the formation of FeBe$_5$ in Be-rich compounds.
Experimental phase diagrams, although tentative and based on limited data \cite{Teitel1948,Okamoto1988}, show a first order transition from $\varepsilon$ to FeBe$_5$ at ${\sim}1450$~K.
On the other hand, the transition between FeBe$_5$ and FeBe$_2$ is not predicted until high temperatures, potentially beyond the melting point of FeBe$_5$.

Secondly, we observe a hexagonal to cubic transition of the FeBe$_2$ phase at high temperature. This is a common feature in many Laves phase systems \cite{Stein2004,Stein2005,Lumley2013,Burr2013a}. Nevertheless, the predicted difference in free energy between the two phases is very small and never exceeds \SI{0.01}{eV/atom}, which is below the level of confidence that the current methodology offers.
Based on these Helmholtz formation energy values convex hull diagrams were created at \SI{0}{\kelvin} (including ZPE) \SI{500}{\kelvin}, \SI{1000}{\kelvin} and \SI{1600}{\kelvin} (see Fig.~\ref{fig:convex_temp}). The FeBe$_2$ phase lies on the convex hull across the entire temperature range. On the other hand, Fe$_2$Be$_{17}$ and FeBe$_5$ are only expected to be stable at low and high temperatures, respectively. At intermediate temperatures the two phases are predicted to co-exist (Fig.~\ref{fig:convex_temp}c).

\begin{figure}[htb]
\centering
	\includegraphics[width=0.45\textwidth]{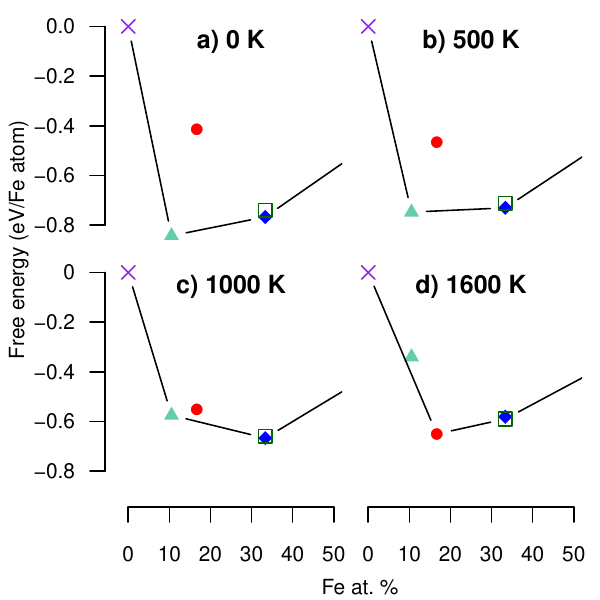}
	\caption{\label{fig:convex_temp} Convex hull diagrams for the Fe-Be system at \SI{0}{\kelvin} (including ZPE), \SI{500}{\kelvin}, \SI{1000}{\kelvin},\SI{1600}{\kelvin}. For the legend see Fig.~\ref{fig:convex}. The values have been normalised as in reactions~\ref{Ef(FeBe12)}--\ref{Esol(Fe_Be)}. For clarity the x-axis was truncated at 50 at.~\% Fe.}
\end{figure}

\subsection{Non-stoichiometry of binary Fe-Be phases}
\label{sec:non-stoich1}

To investigate the accommodation of non-stoichiometry in the intermetallics, the formation enthalpies of intrinsic defects were calculated. In particular, the Fe and Be vacancies, Be substituting for Fe and vice versa. Interstitial Be atoms were also considered.\footnote[3]{Fe interstitial defects were considered unimportant, however, to check this we calculated the energies of Fe$_i$ in FeBe$_2$ and found values typically \SI{ > 5}{eV}, that is, much higher than for equivalent substitutional related process}
 In Kr\"{o}ger-Vink notation these are $V_{\Fe}$, $V_{\Be}$, $\Be_{\Fe}$, $\Fe_{\Be}$, and $\Be_i$, respectively.
\end{alphafootnotes}

Commercial Be alloys are best represented by excess-Be conditions:
in typical alloys, intermetallics only occupy a minute volume fraction, in the form of nano-to-micron sized second phase particles surrounded by metallic Be. Furthermore, as expressed by reactions~\ref{Ef(FeBe2)}--\ref{Esol(Fe_Be)}, at equilibrium, most Fe is expected to be sequestrated within the intermetallic compounds; very little of it is expected to be in solution and none in the form of metallic Fe particles.
In practice this means that there is a readily available reservoir of Be atoms and mass action is achieved by adding or subtracting atoms from bulk Be (reactions~\ref{V_Be}--\ref{Be_i}). The resulting enthalpies correspond to the standard defect enthalpies of formation. On the other hand, the only reservoir of Fe atoms are the intermetallics themselves. Therefore when forming Fe defects, a unit of intermetallic must decompose into free Fe and Be. Fe will react to form the defect and the Be atoms are released into the bulk (reaction~\ref{Fe_Be}). Similarly, defects occupying the Fe site will cause the displaced Fe to react with bulk Be to form one formula unit of the pre-existing intermetallic phase (reactions~\ref{V_Fe+IM}--\ref{Be_Fe+IM}).
The enthalpies of formation of these defects are presented in Table~\ref{tab:non-stoichFeBe}.
\begin{align}
\label{V_Be}		\Be_{\Be}			&\rightarrow	V_{\Be} + \Be\s	\\
\label{Be_i}		\Be\s				&\rightarrow	\Be_i 	\\
\label{Fe_Be}		\Fe\Be_x + \Be_{\Be}	&\rightarrow	\Fe_{\Be} + (x+1) \Be\s	\\
\label{V_Fe+IM} 	\Fe_\Fe + x \Be\s 	& \rightarrow	V_{\Fe} + \Fe\Be_x \\
\label{Be_Fe+IM} 	(x+1) \Be\s + \Fe_\Fe	& \rightarrow	\Be_{\Fe} + \Fe\Be_x
\end{align}
The enthalpies arising from reactions~\ref{Fe_Be}--\ref{Be_Fe+IM}, do not correspond to standard defect formation enthalpies, which requires excess $\Fe\s$. Although not relevant for the current work, the standard formation reactions of those defects and the their energies are reported in \ref{sec:std-Ef}.

\begin{table*}[hbt]
\small
\centering
\caption{
\label{tab:non-stoichFeBe} Formation enthalpy of defects (in eV) that may accommodate non-stoichiometry in FeBe$_2$, FeBe$_5$ and Fe$_2$Be$_{17}$. Defects on Be sites are presented in order of increasing multiplicity. 
Details of the interstitial configurations are presented in the last column.}
\begin{tabular}{l S S | S S S l}
\toprule
	&	\multicolumn{2}{c|}{Be depleted}	&	\multicolumn{3}{c}{Fe depleted}		\\
	&	$\Fe_{\Be}$	&$V_{\Be}$		&$V_{\Fe}$	&$\Be_{\Fe}$	&$\Be_{i}$	&interstitial configurations	\\
\midrule
FeBe$_2$
	&1.88	&1.95	&2.53	&0.30	&5.99	&$1c$					\\
	&2.11	&2.10	&		&		&4.45	&$3j (x=0.488)$			\\
	&		&		&		&		&3.67	&$6m (x=y=0.282)$			\\
	&		&		&		&		&3.52	&$6n (x=\sfrac{1}{2}, z=0.467)$	\\
\midrule
FeBe$_5$
	&-0.67	&2.39	&1.29	&-0.42	&8.29	&$4b$					\\
	&1.07	&1.84	&		&		&7.28	&$4d$					\\
	&		&		&		&		&1.82	&$\textless111\textgreater$ dumbbell on $4c$	\\
	&		&		&		&		&2.31	&$24f (x=0.726)$			\\
\midrule
Fe$_2$Be$_{17}$
	&0.79	&1.56	&1.79	&-0.15	&2.69	&$\textless001\textgreater$ dumbbell on $2a$	\\
	&0.84	&1.48	&		&		&1.74	&$4f$					\\
	&0.89	&1.58	&		&		&1.89	&$6h (x=0.459)$			\\
	&0.90	&2.67	&		&		&2.30	&$12k (x=\sfrac{1}{6}, z=0.938)$\\
	&1.82	&2.02	&		&		&		&	\\					
	&1.88	&1.74	&		&		&		&	\\					
	&2.08	&1.94	&		&		&		&	\\					
\bottomrule
\end{tabular}
\end{table*}

In the case of FeBe$_2$, substitutional and vacancy defects have significantly lower formation enthalpies compared to the interstitial defects.
Defects producing FeBe$_{2+x}$ (reaction~\ref{Be_Fe+IM} proceeding with \SI{0.30}{eV}) are markedly easier to accommodate than those that form FeBe$_{2-x}$.
Vacancy mediated accommodation is markedly less favourable but again the defects that lead to accommodation of excess Be ($V_{\Fe}$) are more stable than those that accommodate excess Fe ($V_{\Be}$). 

In the case of FeBe$_5$, again the lowest energy defects are substitutional, however, in this phase they are negative. This is not surprising considering that FeBe$_5$ was found to be unstable without thermal contributions and should decompose into a combination of Be-rich and Be-poor intermetallics (see sections~\ref{sec:binary_stability} and \ref{sec:thermoFeBe}).
It is, therefore, expected that deviations from stoichiometry are also favourable. For instance, the substitution of Fe onto an FCC-Be site, effectively creates one primitive unit cell of the very stable FeBe$_2$  (C15 polymorph) within the FeBe$_5$ structure.
Experimentally, the solubility range of FeBe$_5$ is recorded to be large (8.33--16.55 at.~\%~\cite{Janot1971}). Here we propose that this is achieved by a substitutional mechanism on both sides of the stoichiometric composition (i.e.\ FeBe$_{5-x}$ and FeBe$_{5+x}$).

Regarding the $\varepsilon$ phase, the work by Johnson \etal \cite{Johnson1970} suggests that this phase may exhibit partial occupancy. The results from Table~\ref{tab:non-stoichFeBe}, suggest that vacancies of either Fe and Be atoms are energetically unfavourable. However, the $\Be_{\Fe}$ defect exhibits negative formation energy. The presence of defects of this type would reduce the Fe content of the compound from 10.5 at.~\% for the stoichiometric Fe$_2$Be$_{17}$ to a value closer to the observed 8 at.~\% value. Therefore, we propose that $\varepsilon$ phase is best represented by the chemical formula Fe$_{2-x}$Be$_{17+x}$, where $x \sim 0.48$.

\section{Ternary Al-Fe-Be phase}
\subsection{Formation of AlFeBe$_4$}
\label{sec:Al_additions}
The binary Al-Be system exhibits no intermetallic phases, and the mutual solid solubilities (Be in Al and Al in Be) are very limited \cite{Okamoto2006b}. A binary Al-Be alloy would therefore only contain single element phases of HCP-Be and FCC-Al.
A ternary Al-Fe-Be compound with high Be content has been reported previously, with composition AlFeBe$4$\cite{Carrabine1963,Rooksby1962}.
A recent DFT study \cite{Middleburgh2011}, showed that in the presence of Fe, Al can react to form  AlFeBe$_4$ following reactions~\ref{Al+Be+Fe} or \ref{Al+FeBe5}.
\begin{align}
\label{Al+Be+Fe}	\Al + 4\Be + \Fe	&\xrightarrow{\SI{-1.30}{eV}} 	\AlFeBe_4	\\
\label{Al+FeBe5}	\Al + \FeBe_5			&\xrightarrow{\SI{-0.85}{eV}} 	\AlFeBe_4	+ \Be		
\end{align}
The enthalpies of reaction calculated in the current work (above) are in close agreement with the previous study (\SI{-1.30}{eV} and \SI{-0.85}{eV} vs \SI{-1.42}{eV} and \SI{-0.89}{eV}, respectively).
Following the results of section~\ref{sec:FeBe}, which suggested that FeBe$_2$ and Fe$_2$Be$_{17}$ are the stable phases at low temperatures, the reactions between Al and these phases were also found to be exothermic (reaction~\ref{Al+FeBe2} and \ref{Al+FBe}):
\begin{align}
\label{Al+FeBe2}	\Al + \FeBe_2 + 2\Be			&\xrightarrow{\SI{-0.49}{eV}} 	\AlFeBe_4 \\
\label{Al+FBe}		\Al + \tfrac{1}{2} \Fe_2\Be_{17} 	&\xrightarrow{\SI{-0.33}{eV}} 	\AlFeBe_4 + \tfrac{9}{2} \Be
\end{align}
The implications are that in the presence of excess Al, the ternary phase is thermodynamically stable. Other ternary phases could theoretically be more stable, however, there is no experimental evidence of other Be-rich ternary compounds, therefore their existence has been discounted.
Magnetic properties of this phase were calculated and are reported in \ref{sec:mag}. It was found that AlFeBe$_4$ exhibits significantly less pronounced ferromagnetism compared to the Fe-Be binary intermetallics.

\subsection{Accommodation of dilute Al additions in the Fe-Be system}
\label{sec:diluteAl}
The incorporation of Al as a dilute point defect into binary Fe-Be intermetallic phases was investigated to model dilute Al-content conditions. 
Since the addition of Al may act as a stabilising agent for some of the metastable intermetallic phases, all binary Fe-Be phases were considered.
Al atoms (calculated metallic radius $r_{Al} = 1.425$ \si{\angstrom}) are significantly larger than Be and Fe atoms ($r_{Be} = 1.109$ \si{\angstrom}, $r_{Fe} = 1.238$ \si{\angstrom}) and therefore unlikely to occupy interstitial sites. To test this, the defect energy of Al in the largest interstitial site was determined, and found to be \SI{5}{eV} less favourable than substitutional defect energies. Instead, substitution onto each of the symmetrically unique Be sites (reaction~\ref{Al_Be}) and the Fe site (reaction~\ref{Al_Fe+IM}) were considered.
Once again, we are interested in the Be-excess conditions. The reactions governing the solution of Al into the intermetallics are:
\begin{align}
\label{Al_Be}		\Al\s + \Be_{\Be}			&\rightarrow	\Al_{\Be} + \Be\s	\\
%\label{Al_Fe(s)}	\Al\s + \Fe_\Fe		 & \rightarrow	\Al_{\Fe} + \Fe\s 	\\
 \label{Al_Fe+IM} 	\Al\s + \Fe_\Fe + x \Be\s 	& \rightarrow	\Al_{\Fe} + \Fe\Be_x 
%\label{Al_Betet}	\Al\s + 4 \left\{\Be_{\Be}\right\} 	&\rightarrow	\Al_{4\Be} + 4\Be\s 	
\end{align}
The standard formation enthalpy (relevant if excess Fe and Al are present) are presented in \ref{sec:std-Ef}.
The solution enthalpies from reactions~\ref{Al_Be} and  \ref{Al_Fe+IM} are reported in Table~\ref{tab:Al_defect_in_FeBeIM}, together with the formation enthalpy of ternary AlFeBe$_4$ (following reactions~\ref{Al+Be+Fe}--\ref{Al+FBe}).

\begin{table}[htb]
\small
\centering
\caption{\label{tab:Al_defect_in_FeBeIM} Solution enthalpy of Al into Fe-Be binary phases together with the formation enthalpy, $E_f$, of AlFeBe$_4$. 
Defects on Be sites are presented in order of increasing multiplicity. %For FeBe$_{12}$, the Be sites are: $8f$, $8i$ and $8j$.
All values in eV.}
\begin{tabular}{l S S |S}
\toprule
	&$\Al_{\Fe}$	&$\Al_{\Be}$	&\multicolumn{1}{c}{$E_f(\AlFeBe_{4})$}\\
\midrule
FeBe$_2$
				&-0.31	&0.79	&-0.49	\\
				&		&0.95	&		\\	\midrule
FeBe$_5$
				&-0.66	&-0.73	&-0.85	\\
				&		&0.94	&		\\	\midrule
Fe$_2$Be$_{17}$
				&-0.07	&0.08	&-0.33	\\
				&		&0.09	&		\\
				&		&0.27	&		\\
				&		&-0.50	&		\\
				&		&0.84	&		\\	
				&		&1.09	&		\\	
				&		&1.26	&		\\	\midrule
FeBe$_{12}$
				&1.04	&1.94	&-1.00	\\
				&		&0.77	&		\\
				&		&1.21	&		\\	\bottomrule
\end{tabular}
\end{table}

Comparing the solution enthalpy with the formation enthalpy of AlFeBe$_4$, it is clear that AlFeBe$_4$ is preferentially formed over dilute Be-Fe-Al ternary intermetallics if sufficient Al is present.
Nevertheless, the dilute incorporation of Al in FeBe$_2$ and FeBe$_5$ yields large and negative solution enthalpies, therefore a degree of solid solution is expected.
On the other hand, the solution of Al into the Be-rich phases is highly unfavourable, in agreement with experiment \cite{Rooksby1962,Myers1978}.

The solution energies in Table~\ref{tab:Al_defect_in_FeBeIM} show that solution of Al in FeBe$_5$ is significantly more favourable than in FeBe$_2$. This suggests that Al may stabilise FeBe$_5$.
Interestingly, the preferred site for Al accommodation in FeBe$_5$ is the FCC-Be ($2a$ Wyckoff site), which is nominally occupied by Al in the AlFeBe$_4$ phase. Accommodation on the FCC-Be site is more favourable than the Fe site, suggesting that the (Fe,Al)Be$_5$ ternary phase originally predicted by Rooksby \cite{Rooksby1962}, is unlikely to form. This agrees with the work of Carrabine \cite{Carrabine1963}.
If the incorporation of Al onto the FCC-Be site continued (unchanged) until $\Al/\Fe=1$, then the AlFeBe$_4$ phase would be formed.
However, the accommodation energy onto the Fe site is only \SI{0.07}{eV} more positive, therefore, as the reaction progresses, it is expected that some of the Al will occupy the Fe site and some the FCC-Be site. The combined reactions, together with the fact that the displaced Fe will either form one extra formula unit of FeBe$_5$ or substitute for an FCC-Be (see section~\ref{sec:non-stoich1}), leads to the formation of disordered (Al,Fe)$_2$Be$_4$ instead of ordered AlFeBe$_4$. The sparse literature available for the ternary Al-Fe-Be phase is inconclusive regarding order/disorder \cite{Rooksby1962,Rooksby1962a,Carrabine1963}.

Regarding FeBe$_2$, the only favourable solution energy is found for substitutions onto the Fe site. Similarly to the previous case, as the incorporation reaction progresses, the host intermetallic FeBe$_2$ tends to become disordered (Al,Fe)Be$_2$, or (Al,Fe)$_2$Be$_4$. Therefore, in the presence of Al, both FeBe$_2$ and FeBe$_5$ will react with any Al in the system and tend towards (Al,Fe)$_2$Be$_4$.

%There is a consideration to be made: the calculated solution energies are strictly valid only at the dilute limit. This limits our scope of prediction to small $\Al/\Fe$ ratios. Nevertheless, the change in solution energy with Al concentration is likely to be a smoothly varying function, therefore it is reasonable to speculate that since the solution energy at the dilute limit is favourable, and the formation energy of AlFeBe$_4$ is favourable (reactions~\ref{Al+Be+Fe}--\ref{Al+FBe}), then the incorporation reaction is likely to be favourable at intermediate compositions.

\section{Order/disorder in the intermetallic phases}
\label{sec:disorder}

\begin{figure}[htb]
	\centering
	\includegraphics[width=0.33\textwidth]{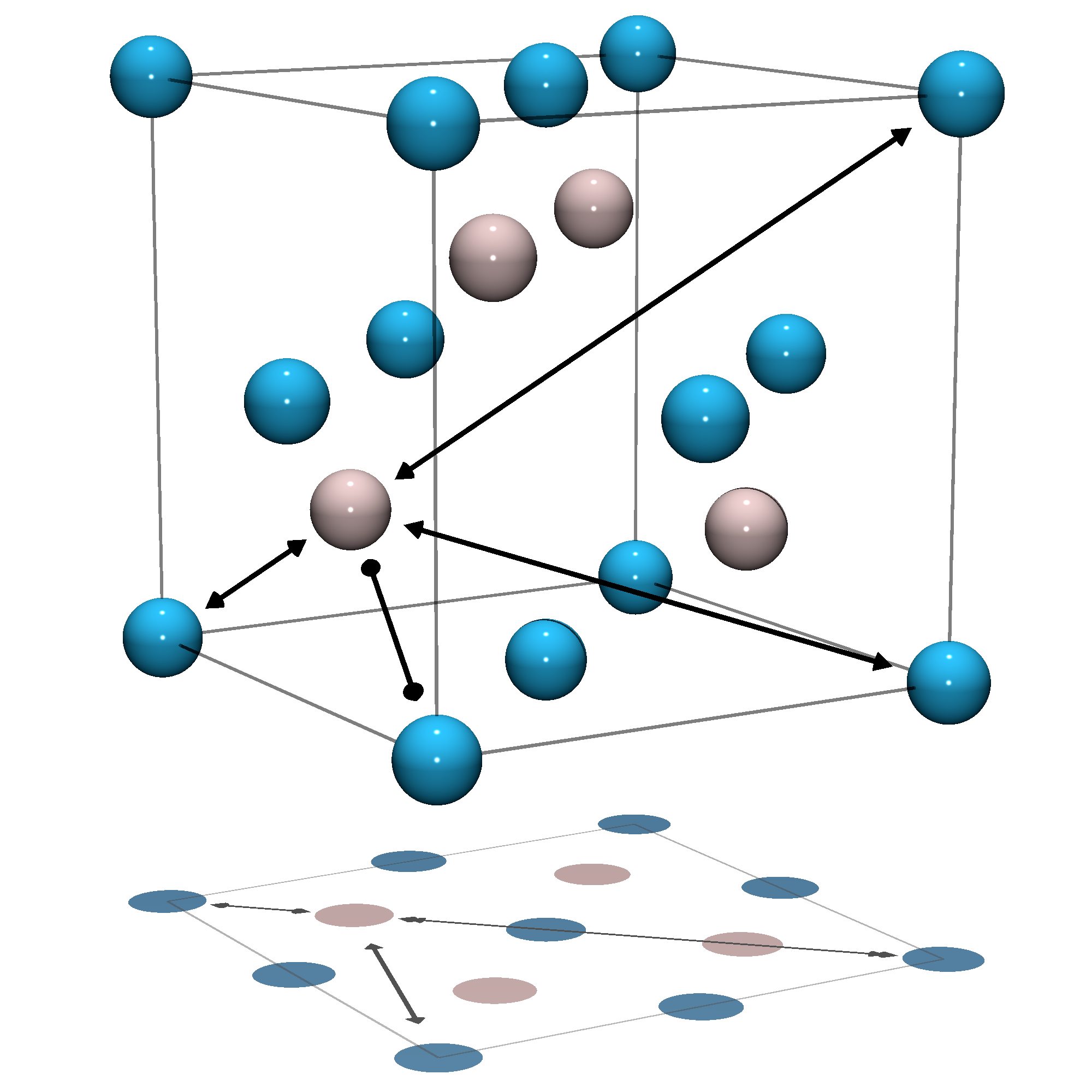}
	\caption{\label{fig:Antisite} Antisite configurations (up to 4nn) in a unit-cell of AlFeBe$_4$, where the Be tetrahedra have been removed for clarity. The two FCC sublattices (blue and light pink) are equivalent.}
\end{figure}

We investigated the driving force for ordering by computing antisite defect energies.
Both dilute (non-interacting) and bound antisite defect pairs were studied. In a $2\times2\times2$ supercell of AlFeBe$_4$, containing 198 atoms, bound antisite pairs on the Fe and Al sublattices can be investigated at separations from \SI{2.55}{\angstrom}, as the first nearest neighbour (1nn), to \SI{7.66}{\angstrom} (4nn) (see Fig.~\ref{fig:Antisite}).
Equivalent simulations were carried out for FeBe$_2$, FeBe$_5$, where only FCC-Be were considered for antisite pairs as these most easily accommodate Fe atoms (see Table~\ref{tab:non-stoichFeBe}).
The results are presented in Table~\ref{tab:antisites}.

\begin{table*}[htb]
\small
\centering
\caption{\label{tab:antisites} Formation energy of antisite defects in binary and ternary intermetallic phases of Be-Fe(-Al). All values in eV.}
\begin{tabular}{l c S S S S S}
\toprule
Phase	&		defect pair& 
\text{1nn}&  \text{2nn}&  \text{3nn}&  \text{4nn}&  \text{Unbound}\\
\midrule
AlFeBe$_4$		&Fe$_{\text{Al}}$-Al$_{\text{Fe}}$		&-0.08	&-0.05	&-0.07	&-0.11	&0.00\\
FeBe$_2$			&Fe$_{\text{Be(2a)}}$-Be$_{\text{Fe}}$	&1.99	&2.13	&2.22	&2.14	&2.18\\
FeBe$_5$			&Fe$_{\text{Be(4c)}}$-Be$_{\text{Fe}}$	&0.67	&0.65	&0.64	&0.66	&-1.09\\
%Fe$_2$Be$_{17}$	&Fe$_{\text{Be}}$-Be$_{\text{Fe}}$		&0.80	&1.91	&-0.01	&1.70	&0.68	\\
%				&								&1.66	&1.71	&1.66	&1.71	&	\\
\bottomrule
\end{tabular}
\end{table*}

For the ternary phase, the defect formation energies (see Table~\ref{tab:antisites}) are negative for the bound defects, and zero for the dilute case. This is a strong indication that the ordered AlFeBe$_4$ as proposed by \cite{Rooksby1962} is unstable and that there is no driving force for ordering in this phase. Therefore, we expect the Al-Fe baring intermetallic phases of Be to exhibit the (Al,Fe)$_2$Be$_4$ Laves structure, where the two FCC sublattices are indistinguishable. 
This agrees with Rooksby \cite{Rooksby1962}, but applied to the correct stoichiometry reported by Carrabine \cite{Carrabine1963} and Myers \etal \cite{Myers1978}.

A competing contribution to the disorder of the phase is the ferromagnetic behaviour found in AlFeBe$_4$, discussed in~\ref{sec:mag}.
Whilst the disordered material may not have any long range magnetic ordering, it may still maintain some local spin polarisation around the Fe atoms and/or  clusters of spin polarised material surrounded by non spin-polarised species~\cite{Middleburgh2014}. To quantify the contributions of magnetic moments to the driving force for ordering, the difference between FM and non-magnetic (NM) configurations provide the upper bound: this is calculated to be \SI{0.16}{eV} per unit cell. This is commensurate with the defect formation energy of a single antisite pair and is therefore not sufficient to promote an ordered structure.

Regarding FeBe$_5$, all bound configurations exhibit a small yet positive defect formation energy, suggesting that the defect concentration will be temperature dependant (i.e.\ the ground state phase is ordered) although given the small energy significant disorder may be anticipated. This is a often an indication of radiation tolerance in the material \cite{Sickafus2000b,Rushton2007a,Sickafus2007a}.
Conversely, the dilute antisite pair in FeBe$_5$ (which is evaluated by considering the effect of accommodating $\Fe_{\Be}$ and $\Be_{\Fe}$ in two spatially separated sites with no interaction between them) has a strongly negative formation energy, which is related to the predicted instability of the phase at low temperatures and its ability to accommodate non-stoichiometry (see section~\ref{sec:FeBe}). 

\begin{figure}[htb]
	\centering
	\includegraphics[width=0.49\textwidth]{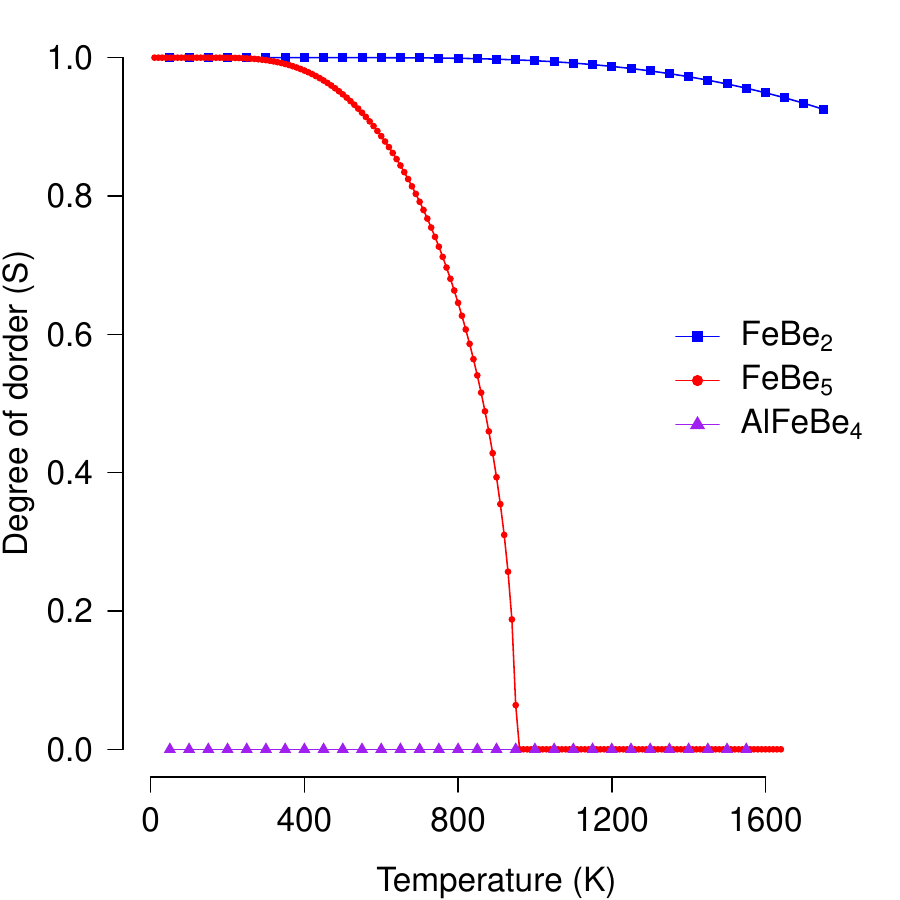}
	\caption{\label{fig:order} Degree of order as a function of temperature for FeBe$_2$ (blue squares), FeBe$_5$ (red circles) and AlFeBe$_4$ (purple triangles).}
\end{figure}

Employing the Bragg-Williams approach, the degree of order in FeBe$_2$, FeBe$_5$ and AlFeBe$_4$ intermetallics was predicted as a function of temperature between \SI{0}{K} and \SI{1500}{K} (see Fig.~\ref{fig:order}).
The Al-baring compound exhibits no order across the entire temperature range, due to the zero formation energy for antisite pairs. In contrast FeBe$_2$ exhibits a high degree of order up to its melting point, while FeBe$_5$ though ordered at low temperatures, exhibits a decrease in ordering at ${\sim}700$~K and complete disorder at temperatures above \SI{950}{K}.
The degree of order is calculated regardless of the stability of the phases. For instance, FeBe$_5$ is predicted to be unstable at temperatures below $\sim1250$~K temperatures (see section~\ref{sec:binary_stability}--\ref{sec:thermoFeBe}), therefore the predicted degree of order is not relevant unless the phase is first stabilised at low temperature (e.g.~ by accommodating low quantities of Al impurities).
These predictions could be tested experimentally through measurements of the specific heat, since a spike in specific heat should be observed in the vicinity of the critical temperature for ordering \cite{Bragg1935}.

%%%%%%%%%%%%%    Conclusions    %%%%%%%%%%%%%
\section{Summary}
\label{sec:conc}
Be is the plasma facing material of choice in current fusion reactor designs.  Fe and Al are common elements found in Be, either as alloying additions or as unintentional impurities and their influence on the performance of Be will depend upon which phases are manifest.  While it is well established that Be rich intermetallics are formed in the presence of Fe and Al, there is conflicting experimental data.  Here we have used atomic scale quantum mechanical simulations based on density functional theory to provide data that we use to predict the structures and energies of various intermetallics and compare these with the solution energies of Fe and Al in Be metal.  While previous simulations have focused on enthalpies alone, here by calculating the phonon DOS of the various phases, temperature effects are included by determining both vibrational enthalpy and entropy contributions --- and hence we base our discussions around the Helmholtz free energy.
The information gathered here provides the foundation for further thermodynamical modelling (e.g.\ using {\sc calphad}) \cite{Kormann2014}.

A commonly observed intermetallic is the so-called $\varepsilon$ phase, however, its stoichiometry and structure are not well established.  Of the 13 candidate considered in this study, a Fe$_2$Be$_{17}$ phase, exhibiting a hexagonal RhBe$_{6.6}$ structure with space group $P\bar{6}m2$ is identified as the most likely, with potential Fe deficiency as highlighted below.
Fe$_2$Be$_{17}$ is found to be stable only at temperatures below ${\sim}1500$~K, beyond which it decomposes into solid solutions of FeBe$_5$ and Be(s).
The intermetallics FeBe$_2$ and FeBe$_5$ are better characterised experimentally although we find that FeBe$_5$ is unstable at temperatures below  ${\sim}1100$~K, unless small additions of Al are present, which are found to stabilise FeBe$_5$ phase above other binary phases. Furthermore, FeBe$_5$ is disordered at the temperatures in which it is stable, and may only becomes fully ordered below \SI{500}{K}, whereas FeBe$_2$ exhibits little disorder until at least \SI{1500}{K}.
%The intermetallics FeBe$_2$ and FeBe$_5$ are better characterised experimentally although we find that FeBe$_5$ is disordered at the temperatures in which it is stable, and only becomes fully ordered below \SI{500}{K}, whereas FeBe$_2$ exhibits little disorder until at least \SI{1500}{K}.
%In terms of phase stability, convex hull diagrams indicate that below ${\sim}1250$~K only FeBe$_2$ and Fe$_2$Be$_{17}$ are predicted to form at equilibrium; above ${\sim}1650$~K only FeBe$_2$ and FeBe$_5$  are observed; and at intermediate temperature all three phases co-exist.
%until \SI{\sim1450}{K} only FeBe$_2$ and Fe$_2$Be$_{17}$ are observed at equilibrium but then FeBe$_5$ replaces the $\varepsilon$ phase.  

Point defects are calculated for all the phases in order to identify the likely extent of deviations from stoichiometric compositions.  FeBe$_5$ exhibits considerable non-stoichiometry with both Fe and Be excess compositions.  Conversely, in Fe$_2$Be$_{17}$ substitution of Be for some Fe is energetically favourable and thus Fe$_2$Be$_{17}$ will be Fe deficient, while defects in FeBe$_2$ are high in energy and this phase will remain much more stoichiometric.

While the binary Al-Be system exhibits no intermetallic phases, Al is readily incorporated into Fe-Be intermetallics. In particular, Al substitution for Be in FeBe$_5$, and for Fe and BE in FeBe$_2$ lead to the AlFeBe$_4$ phase. Disorder is also apparent in this ternary system with no driving force for ordering so that AlFeBe$_4$ should more accurately be reported as (Al,Fe)$_2$Be$_4$, where the Fe and Al sublattices are indistinguishable.

%%%%%%%%%%%    Acknowledgements    %%%%%%%%%%%
\section{Acknowledgements}
We would like to acknowledge the EPSRC and ANSTO for financial support. The computing resources were provided by Imperial College London HPC, and the MASSIVE cluster at Melbourne. Lyndon Edwards is acknowledged for his continuing support, Samuel T.~Murphy and Michael W.D.~Cooper for the fruitful discussions and Simon C.~Lumley for his insight on lattice dynamics.

\appendix

\section{Magnetism in the intermetallics}
\label{sec:mag}
FeBe$_2$, FeBe$_5$, Fe$_2$Be$_{17}$ and AlFeBe$_4$ may exhibit one of a range of magnetic orders, as summarised in Table~\ref{tab:SpinState}. In all cases ferromagnetic (FM) ordering is the most favourable, followed by the high spin antiferromagnetic configuration (AFM-high) in FeBe$_5$ and AlFeBe$_4$ (although the ordered AlFeBe$_4$ structure is not predicted to form).
%Conversely, FeBe$_{12}$ relaxed only to the non-magnetic configuration, while the Fe$_2$Be$_{17}$ phase only shows a slight preference for magnetic ordering, indicating that the order may be lost at high temperatures. 
For all materials under investigation a magnetic transition is expected, and further work may be carried out to calculate the Curie transition using the  \emph{ab-initio} techniques thoroughly reviewed in \cite{Kormann2014}, or by means of heat capacity measurements.

\begin{table}[htb]
\small
\centering
\caption{\label{tab:SpinState}  Energy difference between non-spin polarised calculations (NM), and possible stable magnetic configurations. Values are reported in eV and normalised per conventional unit cells.}
\begin{tabular}{l S S S S S S}
\toprule
	&	\text{NM}	&\text{AFM-low}	&\text{AFM-high}	&\text{FM}		\\	
\midrule
$\Fe_{2}\Be_{17}$	&0.00		&\text{---}	&\text{---}	&-0.04	\\
$\FeBe_{5}$		&0.00		&-0.49	&-0.59	&-0.67	\\
$\FeBe_{2}$		&0.00		&-0.01	&-1.24	&-1.66	\\
$\AlFeBe_{4}$		&0.00		&-0.08	&-0.14	&-0.16	\\
\bottomrule
\end{tabular}
\end{table}

The reported energy differences correspond to a conventional unit cell. 
This non-negligible contribution to the energy of the system should be considered when computing isolated defects. In such simulations, the presence of a defect may cause the minimisation algorithm to converge into in a shallow minima with a metastable spin state. In the current work, no constrains were added to the spin while performing an energy relaxation to allow localised changes of the spin near a defect, but great care was taken to ensure that the overall spin state of the system was unchanged after the introduction of a defect. When that did not occur, the simulations were restarted with a slightly different initial spin state and tighter electronic convergence criteria, to help the minimiser overcome local barriers and find the lowest energy minimum.
In all cases, it was found that the non-ferromagnetic solution was not the lowest energy configuration.

Regarding the ordered AlFeBe$_4$ phase, it exhibits similar, yet significantly less pronounced, magnetic properties compared to FeBe$_5$ phases. Long range magnetic ordering, both FM and AFM, are not maintained when Fe and Al atoms are randomly distributed in the FCC sublattices. As the ordered AlFeBe$_4$ structure was found to be unstable, preferring to form disordered (Al,Fe)$_2$Be$_4$, the ternary compound is expected to exhibit no magnetic ordering (see section~\ref{sec:disorder}).

\section{Elastic constants}
\label{sec:elastics}

The complete stiffness matrices were calculated for all the intermetallic phases in the Fe-Be system (see Table~\ref{tab:elastics}).
These were obtained by performing small lattice perturbations from the ground state structures, and measuring the stresses, while keeping all relative atomic positions fixed. Bulk moduli ($K$) and shear moduli ($G$) were evaluated using the Voigt-Reuss-Hill method (Hill average) \cite{Hill1952a}.

\begin{table*}[hbt]
\small
\centering
\caption{\label{tab:elastics} Ground state elastic constants of binary intermetallic phases. In all cases $C_{11} = C_{22}$, $C_{13} = C_{23}$ and $C_{44} = C_{55}$. All values are expressed in units of GPa. Uncertainties are below \SI{1.5}{\%} unless otherwise stated.}
\begin{tabular}{l S S S S S S S S}
\toprule
Phase		&	$C_{11}$&	$C_{33}$&	$C_{12}$&	$C_{13}$&	$C_{44}$&	$C_{66}$ &	$K$	& 	$G$\\
\midrule
FeBe$_{12}$&		345.0&	319\pm7&	20\pm1&	41\pm2&	131.7&	109.2&	134.7&	133.8	\\
Fe$_2$Be$_{17}$&	318.6&	377.8&	66.8&	25.8&	107.0&	125.9&	139.1&	125.8	\\
FeBe$_5$	&		285.2&	285.2&	69.3&	69.3&	138.6&	138.6&	141.3&	125.4	\\
FeBe$_2$	&		361.7&	378.3&	61.4&	57.3&	160.2&	150.2&	161.5&	155.8	\\
%AlFeBe$_4$&		349.8&	349.8&	54.2&	54.2&	142.2&	142.2&	152.7&	144.4\\
\bottomrule
\end{tabular}
\end{table*}

\section{Defect formation energy from standard state}
\label{sec:std-Ef}

\begin{table}[hbt]
\small
\centering
\caption{\label{tab:non-stoich_excessFe} \emph{Standard} formation enthalpy of defects involving Fe atoms in FeBe$_2$, FeBe$_5$ and Fe$_2$Be$_{17}$.}
\begin{tabular}{l|S S S S}
\toprule
Phase	&	$V_{\Fe}$\text{(\ref{V_Fe(s)})}&	$\Be_{\Fe}$\text{(\ref{Be_Fe(s)})}&	$\Fe_{\Be}$\text{(\ref{Fe_Be(s)})}&$\Al_{\Fe}\text{(\ref{Al_Fe(s)})}$\\
\midrule
%FeBe$_{12}$&	&	&	&	&1.34 	\\
\multirow{3}{*}{Fe$_2$Be$_{17}$}
		&2.76	&0.82	&-0.19	&0.90\\
		&		&		&-0.14	&	\\
		&		&		&-0.08	&	\\
		&		&		&-0.08	&	\\
		&		&		&0.84	&	\\
		&		&		&0.90	&	\\
		&		&		&1.10	&	\\	\midrule
FeBe$_5$	&1.74	&0.02	&-1.11	&-0.22\\
		&		&		&0.63	&	\\	\midrule
FeBe$_2$	&3.34	&1.11	&1.07	&0.50\\	
		&		&		&1.30	&	\\
	
\bottomrule
\end{tabular}
\end{table}

The standard defect formation enthalpy of intrinsic Fe defects are calculated following reactions~\ref{V_Fe(s)}--\ref{Al_Fe(s)}, and the results are presented in Table~\ref{tab:non-stoich_excessFe}.
\begin{align}
\label{V_Fe(s)}		\Fe_\Fe		 	& \rightarrow	V_{\Fe} + \Fe\s	\\
\label{Be_Fe(s)}	\Fe_\Fe	+ \Be\s	& \rightarrow	\Be_{\Fe} + \Fe\s	\\
\label{Fe_Be(s)}	\Fe\s + \Be_{\Be}	&\rightarrow	\Fe_{\Be} + \Be\s	\\
\label{Al_Fe(s)}	\Al\s + \Fe_\Fe		 & \rightarrow	\Al_{\Fe} + \Fe\s 
\end{align}

These reactions are likely to occur only if $\Fe\s$ is present at equilibrium, a situation not found in commercial Be alloys.
These enthalpies of formation are significantly less favourable compared to those in Table~\ref{tab:non-stoichFeBe}. This indicates that the formation of one extra formula unit of the existing intermetallics (as per reactions~\ref{Fe_Be}--\ref{Be_Fe+IM}), greatly reduces the energy penally for accommodation of substitutional defects in Fe-Be binary intermtallics.

%%%%%%%%%%%%%%%%%       BIBLIOGRAPHY        %%%%%%%%%%%%%%%%%%%%%

\section*{References}

%%\bibliographystyle{elsarticle-num.bst}
%%\bibliographystyle{model1a-num-names}
%\bibliographystyle{../pre_submission/elsarticle-num-names1.bst}

%%\bibliography{/Users/pab07/Documents/papers/library}			%auto-sync
%\bibliography{/Users/pab07/Documents/papers/collection}

%%%%%%%%%%%%%%%%              the bbl

\clearpage

\end{document}